\documentclass[12pt]{article}
 
\usepackage{here}
\usepackage{amsmath,amssymb}
\usepackage{bbold}
\usepackage{bm}
\usepackage[dvipdfmx]{graphicx}
\usepackage{graphicx, color}
\usepackage{wrapfig}
\usepackage{cite}
\usepackage{here}
\begin{document}
\title{
\begin{flushright}
\ \\*[-115pt] 
\begin{minipage}{0.22\linewidth}
\normalsize
APCTP Pre 2021-009\\
HUPD-2103\\
 \\*[35pt]
\end{minipage}
\end{flushright}
{\Large \bf 
 Modulus $\tau$ linking leptonic CP violation 
 to baryon \\ asymmetry 	in  $A_4$ modular invariant flavor  model
\\*[15pt]}}

\author{ 
\centerline{
Hiroshi Okada $^{a,b}\footnote{E-mail address: hiroshi.okada@apctp.org}$,
~~Yusuke Shimizu $^{c,d}\footnote{E-mail address: yu-shimizu@hiroshima-u.ac.jp}$,
~~Morimitsu Tanimoto $^{e}\footnote{E-mail address: tanimoto@muse.sc.niigata-u.ac.jp}$} \\*[5pt]
  \centerline{and \quad Takahiro Yoshida $^{e}\footnote{E-mail address: yoshida@muse.sc.niigata-u.ac.jp}$} \\*[5pt]
\centerline{
\begin{minipage}{\linewidth}
\begin{center}
$^a${\it \normalsize
Asia Pacific Center for Theoretical Physics, Pohang 37673, Republic of Korea} \\*[5pt]
$^b${\it \normalsize
Department of Physics, Pohang University of Science and Technology, Pohang 37673,\\ Republic of Korea} \\*[5pt]
$^c${\it \normalsize
	Physics Program, Graduate School of Advanced Science and Engineering,
	Hiroshima University, Higashi-Hiroshima 739-8526, Japan} \\*[5pt]
$^d${\it \normalsize
Core of Research for the Energetic Universe, Hiroshima University,
Higashi-Hiroshima 739-8526, Japan} \\*[5pt]
$^e${\it \normalsize
Department of Physics, Niigata University, Niigata 950-2181, Japan}
\end{center}
\end{minipage}}
\\*[90pt]}

\date{
\centerline{\small \bf Abstract}
\begin{minipage}{0.9\linewidth}
\medskip 
\medskip 
\small 
We propose an $A_4$  modular invariant flavor model of leptons,
in which  
both CP and modular symmetries are broken spontaneously by the vacuum expectation value  of the modulus $\tau$. 
The value of the modulus $\tau$
 is restricted by
 the observed lepton mixing angles and lepton masses for the normal hierarchy of neutrino masses.
The predictive  Dirac CP phase $\delta_{CP}$
is  in the ranges
 $[0^\circ,50^\circ]$, $[170^\circ,175^\circ]$ and
$[280^\circ,360^\circ]$ for ${\rm Re}\,[\tau]<0$,
and 
$[0^\circ,80^\circ]$, $[185^\circ,190^\circ]$ and
$[310^\circ,360^\circ]$ for ${\rm Re}\,[\tau]>0$.
The sum of three neutrino masses is predicted in  $[60,\,84]$\,meV,
and  the effective mass 
 for the $0\nu\beta\beta$ decay is in [0.003,\,3]\,meV.
The modulus $\tau$ links the Dirac CP phase  
to the cosmological baryon asymmetry (BAU)  via the leptogenesis. 
Due to the strong wash-out effect,   the predictive baryon asymmetry $Y_B$  can be at most the same order of the observed value.
Then, the lightest right-handed neutrino mass
is restricted in the range of
$M_1 =[1.5,\,6.5]  \times 10^{13}$\,GeV.
We find the correlation
between  the predictive $Y_B$ and the   Dirac CP phase $\delta_{CP}$.
Only two predictive $\delta_{CP}$ ranges,   $[5^\circ,40^\circ]$
(${\rm Re}\,[\tau]>0$)
and $[320^\circ,355^\circ]$  (${\rm Re}\,[\tau]<0$)
are consistent with the  BAU. 
\end{minipage}
}

\begin{titlepage}
\maketitle
\thispagestyle{empty}
\end{titlepage}

\section{Introduction}

One interesting approach to the origin
of flavor structure is to impose a flavor symmetry on a theory. 
The non-Abelian discrete groups are attractive  ones to understand
flavor structure of quarks and leptons.
The  $S_3$ flavor symmetry was studied  to understand the large mixing angle
 \cite{Fukugita:1998vn} in the oscillation of atmospheric neutrinos \cite{Fukuda:1998mi} as well as discussing the  Cabibbo angle
  \cite{Pakvasa:1977in, Wilczek:1977uh}.
 For the last twenty years, the  non-Abelian discrete symmetries of flavors have been developed
  \cite{Altarelli:2010gt,Ishimori:2010au,Ishimori:2012zz,Hernandez:2012ra,King:2013eh,King:2014nza,Tanimoto:2015nfa,King:2017guk,Petcov:2017ggy,Feruglio:2019ktm}, that is
  motivated by the precise observation of  flavor mixing angles of  leptons.
Among them, the $A_4$ flavor symmetry provides 
a simple explanation of the  
existence of  three families of quarks and leptons 
\cite{Ma:2001dn,Babu:2002dz,Altarelli:2005yp,Altarelli:2005yx,
Shimizu:2011xg,Petcov:2018snn,Kang:2018txu}.
However,  it is difficult to obtain clear clues of the $A_4$ flavor symmetry because of a lot of free parameters associated with  scalar flavon fields.

An interesting  approach to the lepton flavor problem
has been put forward 
based on the invariance under the modular transformation \cite{Feruglio:2017spp}, 
where the model of the finite
modular group  $\Gamma_3 \simeq A_4$ has been presented.
In this approach, fermion matrices are written in terms of modular forms which are   holomorphic functions of  the modulus  $\tau$.
This work  inspired further studies of the modular invariance approach
 to the lepton flavor problem. 

The finite groups $S_3$, $A_4$, $S_4$, and $A_5$
are realized in modular groups \cite{deAdelhartToorop:2011re}.
Modular invariant flavor models have been also proposed on the $\Gamma_2\simeq S_3$ \cite{Kobayashi:2018vbk},
$\Gamma_4 \simeq S_4$ \cite{Penedo:2018nmg} and  
 $\Gamma_5 \simeq A_5$ \cite{Novichkov:2018nkm}.
Phenomenological studies of the lepton flavors have been done
based on  $\rm A_4$ \cite{Criado:2018thu,Kobayashi:2018scp,Ding:2019zxk}, $\rm S_4$ \cite{Novichkov:2018ovf,Kobayashi:2019mna,Wang:2019ovr} and 
$\rm A_5$ \cite{Ding:2019xna}.
A clear prediction of the neutrino mixing angles and the Dirac CP  phase was given in  the  simple lepton mass matrices  with
the  $\rm A_4$ modular symmetry \cite{Kobayashi:2018scp}.
The  Double Covering groups  $\rm T'$~\cite{Liu:2019khw,Chen:2020udk}
and $\rm S_4'$ \cite{Novichkov:2020eep,Liu:2020akv} were also
realized in the modular symmetry.
Furthermore, phenomenological studies have been developed  in many works
\cite{deMedeirosVarzielas:2019cyj,
Asaka:2019vev,Ding:2020msi,Asaka:2020tmo,Behera:2020sfe,Mishra:2020gxg,deAnda:2018ecu,Kobayashi:2019rzp,Novichkov:2018yse,Kobayashi:2018wkl,Okada:2018yrn,Okada:2019uoy,Nomura:2019jxj, Okada:2019xqk,
Kariyazono:2019ehj,Nomura:2019yft,Okada:2019lzv,Nomura:2019lnr,Criado:2019tzk,
King:2019vhv,Gui-JunDing:2019wap,deMedeirosVarzielas:2020kji,Zhang:2019ngf,Nomura:2019xsb,Kobayashi:2019gtp,Lu:2019vgm,Wang:2019xbo,King:2020qaj,Abbas:2020qzc,Okada:2020oxh,Okada:2020dmb,Ding:2020yen,Nomura:2020opk,Nomura:2020cog,Okada:2020rjb,Okada:2020ukr,Nagao:2020azf,Nagao:2020snm,Yao:2020zml,Wang:2020lxk,Abbas:2020vuy, Du:2020ylx, Okada:2020brs,Yao:2020qyy,Feruglio:2021dte,King:2021fhl,Chen:2021zty,Novichkov:2021evw,Kobayashi:2021bgy,Ding:2021zbg,Kuranaga:2021ujd}
while theoretical investigations have been also proceeded \cite{Kobayashi:2018bff,Baur:2019kwi,Kobayashi:2019xvz,Nilles:2020kgo,Nilles:2020nnc,Kikuchi:2020nxn,Kikuchi:2020frp,Ishiguro:2020nuf,Ding:2020zxw,Ishiguro:2020tmo,Hoshiya:2020hki,Kikuchi:2021ogn,Ding:2021iqp,Nilles:2020tdp,Baur:2020jwc,Nilles:2020gvu,Baur:2020yjl,Baur:2021mtl}.

 In order to test the flavor symmetry, the  prediction
 of the  Dirac CP  phase is important.
 The CP transformation is non-trivial if the non-Abelian  discrete flavor symmetry is set in the Yukawa sector of a Lagrangian.
 Then, we should discuss  so-called the  generalized CP symmetry 
 in the flavor space \cite{Ecker:1981wv,Ecker:1983hz,Ecker:1987qp,Neufeld:1987wa,
 	Grimus:1995zi,Grimus:2003yn}.
 The modular invariance has been also studied in the framework of  the generalized CP symmetry \cite{Novichkov:2019sqv,Kobayashi:2019uyt}.
 It provided a significant scheme to predict Dirac and Majorana CP  phases of leptons.
  A viable lepton model was proposed in the modular $A_4$ 
  symmetry \cite{Okada:2020brs}, in which the CP violation is realized
   by fixing $\tau$, that is, the breaking of the modular symmetry.
   Afterward, the systematic search of the viable $A_4$ model was done
 \cite{Yao:2020qyy}.
 
 The CP violation by the modulus $\tau$  raises a  question.
 Is the leptonic CP violation   linked to  the baryon asymmetry of the universe (BAU)?  The BAU is now measured very precisely by the cosmic microwave background radiation \cite{Aghanim:2018eyx}.
  One of the most studied scenarios for
 baryogenesis is the canonical leptogenesis scenario \cite{Fukugita:1986hr}, in which the decays of right-handed neutrinos can
 generate the lepton asymmetry that is partially converted into the baryon asymmetry via the sphaleron process \cite{Kuzmin:1985mm}. 
 The sign of the BAU is controlled by the CP violation pattern in the leptonic sector. In general, the sign of the BAU cannot be predicted uniquely even if the  Dirac and Majorana CP  phases  are determined. This is because there exist additional phases associated with right-handed neutrinos which  decouple from the low energy phenomena if right-handed neutrinos are sufficiently heavy.
 However,  there are non-trivial relations between the properties of right-handed neutrinos  and the low energy observables of neutrinos  in the $A_4$ modular symmetry. Indeed,
  the modulus $\tau$ controls the CP phases of both the left-handed sector of neutrinos and the right-handed one in our scheme.
   Under these situations, it is interesting to investigate the sign and magnitude of the BAU.
 
 In the framework of  the modular symmetry,
  the  BAU  has been  studied in $A_4$ model of leptons,
  where the source of CP violation is a complex parameter
  in the Dirac neutrino mass matrix
  in addition to the modulus $\tau$ \cite{Asaka:2019vev}.
  In our work,
  the origin of the  CP violation is  only in modulus $\tau$,
 therefore, it is also  only the source of  the leptogenesis.  
We present a modular $A_4$ invariant model
  with  the  CP symmetry, where both CP and modular symmetries are broken spontaneously by the vacuum expectation value (VEV) of the modulus $\tau$.  
 We discuss the phenomenological implication of this model, that is 
 the Pontecorvo-Maki-Nakagawa-Sakata (PMNS) mixing angles \cite{Maki:1962mu,Pontecorvo:1967fh} 
 and the Dirac  CP  phase  of  leptons,
 which is expected to be observed at T2K and NO$\nu$A experiments \cite{T2K:2020,Adamson:2017gxd}.
 Then,
  we examine a link between  the predictive Dirac  CP phase and  the BAU.



The paper is organized as follows.
In section 2,  we give a brief review on the  CP transformation
 in the modular symmetry. 
In section 3,  we present the CP invariant lepton mass matrix in the $A_4$ modular symmetry.
In section 4, we show the phenomenological implication of lepton mixing  and CP phases.
In section 5, we give the framework of the leptogenesis in our model.
In section 6,  we discuss the link between the predictive Dirac CP  phase and  BAU numerically.
Section 7 is devoted to the summary.
In Appendices A and B, we give  the tensor product  of the $A_4$ group and the modular forms, respectively.
In Appendix C, we show the definition of PMNS matrix elements and  how to obtain the Dirac $CP$ phase, the Majorana phases and the effective mass of the $0\nu\beta\beta$ decay. 
In Appendix D, alternative $A_4$ models and their results are presented.
In Appendix E, we give the relevant formulae of the leptogenesis
 explicitly.

\section{CP transformation in modular symmetry}
\subsection{Generalized CP symmetry}

The CP transformation is non-trivial if the non-Abelian  discrete flavor symmetry $G$ is set in the Yukawa sector of a Lagrangian \cite{Grimus:2003yn,Branco:2011zb}.
Let us consider the  chiral superfields.
The CP is a discrete symmetry which involves both Hermitian conjugation of a chiral superfield $\psi(x)$ and inversion of spatial coordinates,
\begin{equation}
\psi(x) \rightarrow {\bf X}_{\bf r}\overline \psi(x_P) \ ,
\label{gCP}
\end{equation}
where $x_P=(t,-{\bf x})$ and ${\bf X_{r}}$ is a unitary transformation
of $\psi(x)$ in the irreducible representation $\bf r$ of the discrete flavor symmetry $G$.  This transformation is so-called
 the generalized CP transformation.
  If ${\bf X_{r}}$ is the unit matrix,  the CP transformation is  the trivial one. 
This is the case for the continuous flavor symmetry \cite{Branco:2011zb}.
However, in the framework of the non-Abelian discrete family symmetry,
non-trivial choices of ${\bf X_{r}}$ are  possible.
The unbroken CP transformations of ${\bf X_{r}}$ form the group $H_{CP}$.
Then, ${\bf X_{r}}$ must be consistent with the flavor symmetry transformation,
\begin{equation}
\psi(x) \rightarrow {\rho}_{\bf r}(g)\psi(x) \ , \quad g \in G \ ,
\end{equation}
where ${\rho}_{\bf {r}}(g)$ is the representation matrix for $g$
in the irreducible representation $\bf {r}$.

The consistent condition is obtained as follows.
At first,  perform a CP transformation
$\psi(x) \rightarrow {\bf X}_{\bf r}\overline\psi(x_P)$,
then apply a flavor symmetry transformation, 
$\overline\psi(x_P) \rightarrow {\rho}_{\bf r}^*(g)\overline\psi(x_P)$,
and finally perform an inverse CP transformation.
The whole transformation is written as
$\psi(x) \rightarrow {\bf X}_{\bf r} \rho^*(g) 
{\bf X}^{-1}_{\bf r}\psi(x)$,
which must be equivalent to some flavor symmetry
$\psi(x) \rightarrow {\rho}_{\bf r}(g')\psi(x)$. 
Thus, one obtains  \cite{Holthausen:2012dk}
\begin{equation}
{\bf X}_{\bf r} \rho_{\bf r}^*(g) {\bf X}^{-1}_{\bf r}=
{\rho}_{\bf r}(g') 
\ , \qquad g,\, g' \in G \ .
\label{consistency}
\end{equation}
This equation defines the consistency condition, which has to be respected for consistent implementation of a generalized CP symmetry along with a flavor symmetry \cite{Feruglio:2012cw,Chen:2014tpa}.

It has been also shown that  
the full symmetry group is isomorphic to a semi-direct product of $G$ and $H_{CP}$,  that is $G\rtimes H_{CP}$, where  $ H_{CP}\simeq \mathbb{Z}_2^{CP}$
is the group generated by the  generalised CP transformation
under the assumption of $\bf X_{r}$ being a symmetric matrix \cite{Feruglio:2012cw}.

\subsection{Modular symmetry}
The modular group $\bar\Gamma$ is the group of linear fractional transformations
$\gamma$ acting on the modulus  $\tau$, 
belonging to the upper-half complex plane as:
\begin{equation}\label{eq:tau-SL2Z}
\tau \longrightarrow \gamma\tau= \frac{a\tau + b}{c \tau + d}\ ,~~
{\rm where}~~ a,b,c,d \in \mathbb{Z}~~ {\rm and }~~ ad-bc=1, 
~~ {\rm Im} [\tau]>0 ~ ,
\end{equation}
 which is isomorphic to  $PSL(2,\mathbb{Z})=SL(2,\mathbb{Z})/\{\rm I,-I\}$ transformation.
This modular transformation is generated by $S$ and $T$, 
\begin{eqnarray}
S:\tau \longrightarrow -\frac{1}{\tau}\ , \qquad\qquad
T:\tau \longrightarrow \tau + 1\ ,
\label{symmetry}
\end{eqnarray}
which satisfy the following algebraic relations, 
\begin{equation}
S^2 =\mathbb{1}\ , \qquad (ST)^3 =\mathbb{1}\ .
\end{equation}

 We introduce the series of groups $\Gamma(N)$, called principal congruence subgroups, where  $N$ is the level $1,2,3,\dots$.
 These groups are defined by
 \begin{align}
 \begin{aligned}
 \Gamma(N)= \left \{ 
 \begin{pmatrix}
 a & b  \\
 c & d  
 \end{pmatrix} \in SL(2,\mathbb{Z})~ ,
 ~~
 \begin{pmatrix}
  a & b  \\
 c & d  
 \end{pmatrix} =
  \begin{pmatrix}
  1 & 0  \\
  0 & 1  
  \end{pmatrix} ~~({\rm mod} N) \right \}
 \end{aligned} .
 \end{align}
 For $N=2$, we define $\bar\Gamma(2)\equiv \Gamma(2)/\{\rm I,-I\}$.
Since the element $\rm -I$ does not belong to $\Gamma(N)$
  for $N>2$, we have $\bar\Gamma(N)= \Gamma(N)$.
   The quotient groups defined as
   $\Gamma_N\equiv \bar \Gamma/\bar \Gamma(N)$
  are  finite modular groups.
In these finite groups $\Gamma_N$, $T^N=\mathbb{1}$  is imposed.
 The  groups $\Gamma_N$ with $N=2,3,4,5$ are isomorphic to
$S_3$, $A_4$, $S_4$ and $A_5$, respectively \cite{deAdelhartToorop:2011re}.

Modular forms $f_i(\tau)$ of weight $k$ are the holomorphic functions of $\tau$ and transform as
\begin{equation}
f_i(\tau) \longrightarrow (c\tau +d)^k \rho(\gamma)_{ij}f_j( \tau)\, ,
\quad \gamma\in G\, ,
\label{modularforms}
\end{equation}
under the modular symmetry, where
  $\rho(\gamma)_{ij}$ is a unitary matrix under $\Gamma_N$.


Under the modular transformation of Eq.\,(\ref{eq:tau-SL2Z}), chiral superfields $\psi_i$ ($i$ denotes flavors) with weight $-k$
transform as \cite{Ferrara:1989bc},
\begin{equation}
\psi_i\longrightarrow (c\tau +d)^{-k}\rho(\gamma)_{ij}\psi_j\, .
\label{chiralfields}
\end{equation}

 We study global supersymmetric models, e.g., 
minimal supersymmetric extensions of the Standard Model (MSSM).
The superpotential which is built from matter fields and modular forms
is assumed to be modular invariant, i.e., to have 
a vanishing modular weight. For given modular forms 
this can be achieved by assigning appropriate
weights to the matter superfields.

The kinetic terms  are  derived from a K\"ahler potential.
The K\"ahler potential of chiral matter fields $\psi_i$ with the modular weight $-k$ is given simply  by 
\begin{equation}
K^{\rm matter} = \frac{1}{[i(\bar\tau - \tau)]^{k}} \sum_i|\psi_i|^2,
\end{equation}
where the superfield and its scalar component are denoted by the same letter, and  $\bar\tau =\tau^*$ after taking VEV of $\tau$.
The canonical form of the kinetic terms  is obtained by 
changing the normalization of parameters \cite{Kobayashi:2018scp}.
The general K\"ahler potential consistent with the modular symmetry possibly contains additional terms \cite{Chen:2019ewa}. However, we consider only the simplest form of
the K\"ahler potential.

For $\Gamma_3\simeq A_4$, the dimension of the linear space 
${\cal M}_k(\Gamma{(3)})$ 
of modular forms of weight $k$ is $k+1$ \cite{Gunning:1962,Schoeneberg:1974,Koblitz:1984}, i.e., there are three linearly 
independent modular forms of the lowest non-trivial weight $2$,
which form a triplet of the $A_4$ group,
 ${ Y^{(\rm 2)}_{\bf 3}}(\tau)=(Y_1(\tau),\,Y_2(\tau),\, Y_3(\tau))^T$.
These modular forms have been explicitly given \cite{Feruglio:2017spp}  in the  symmetric base of the 
$A_4$ generators  $S$ and $T$ for the triplet representation
 (see Appendix A)  in  Appendix B.


\subsection{CP transformation of the modulus $\tau$}
The CP transformation in the modular symmetry  was  given by using the generalized CP symmetry  \cite{Novichkov:2019sqv}.
We summarize the discussion in Ref.\cite{Novichkov:2019sqv} briefly.
Consider the CP and modular transformation $\gamma$ of the chiral superfield $\psi (x)$ with weight $-k$ assigned to an irreducible unitary representation $\bf r$ of $\Gamma_N$.
The chain $CP\rightarrow \gamma \rightarrow CP^{-1}=\gamma'\in \bar\Gamma$ is expressed as:
  \begin{align}
  \psi(x) &\xrightarrow{\, CP\, } {\bf X}_{\bf r} \overline \psi(x_P)
 \xrightarrow{\ \gamma\ }
 (c\tau^*+d)^{-k} {\bf X}_{\bf r}\, {\rho}_{\bf r}^*(\gamma)\overline\psi(x_P)
 \nonumber\\
 &\xrightarrow{\, CP^{-1}\, }
 (c\tau^*_{CP^{-1}}+d)^{-k} {\bf X}_{\bf r}\, 
 {\rho}_{\bf r}^*(\gamma){\bf X}_{\bf r}^{-1}\psi(x) \,,
\label{chain}
\end{align}
where $\tau_{CP^{-1}}$ is the operation of  $CP^{-1}$  on  $\tau$.
The result of this chain transformation should be equivalent to
a modular transformation $\gamma'$ which maps $\psi(x)$ to $(c'\tau+d')^{-k}  {\rho}_{\bf r}(\gamma')\psi(x)$.
Therefore, one obtains
 \begin{align}
{\bf X}_{\bf r} \rho_{\bf r}^*(\gamma) {\bf X}^{-1}_{\bf r}=
\left (\frac{c'\tau+d'}{c\tau^*_{CP^{-1}}+d}
 \right )^{-k} {\rho}_{\bf r}(\gamma') 
 \, .
\label{consistency2}
\end{align}
Since ${\bf X}_{\bf r}$, $\rho_{\bf r}$ and $\rho_{\bf r'}$ are independent of $\tau$,
the overall coefficient on the right-hand side of Eq.\,(\ref{consistency2})
	has  to be a constant (complex) for non-zero weight  $k$:
	 \begin{align}
	 \frac{c'\tau+d'}{c\tau^*_{CP^{-1}}+d} =\frac{1}{\lambda^*}\, ,
	\label{constant}
	\end{align}
where $|\lambda|=1$ due to the unitarity of 
$\rho_{\bf r}$ and $\rho_{\bf r'}$. 
The values of $\lambda$, $c'$ and $d'$ depend on $\gamma$.

 Taking $\gamma=S$ ($c=1$,\,$d=0$) , and denoting $c'(S)=C$,  $d'(S)=D$
 while keeping  $\lambda(S)=\lambda$, we find
 $\tau=(\lambda\tau^*_{CP^{-1}}-D)/C$ from Eq.\,(\ref{constant}), and consequently,
 \begin{align}
\tau \xrightarrow{\, CP^{-1}\, }
 \tau_{CP^{-1}}=\lambda(C\tau^*+D)\,,  \qquad 
 \tau \xrightarrow{\, CP\, } \tau_{CP}=\frac{1}{C}(\lambda\tau^*-D)\,. 
\label{tauCP1}
\end{align}
Let us act with chain  $CP\rightarrow T \rightarrow CP^{-1}$
on the mudular $\tau$ itself:
 \begin{align}
\tau \xrightarrow{\, CP\, } \tau_{CP}=\frac{1}{C}(\lambda\tau^*-D) 
\xrightarrow{\ T\ } \frac{1}{C}(\lambda(\tau^*+1)-D)
\xrightarrow{\, CP^{-1}\, } \tau+\frac{\lambda}{C} \, .
\label{tauCP2}
\end{align}
The resulting transformation has to be a modular transformation, therefore
$\lambda/C$ is an integer. Since $|\lambda|=1$, we find $|C|=1$ and $\lambda=\pm 1$.
 After choosing the sign of $C$ as $C=\mp 1$ so that ${\rm Im}[\tau_{CP}] >0$,
  the CP transformation of Eq.\,(\ref{tauCP1}) turns to
   \begin{align}
  \tau \xrightarrow{\, CP\, } n-\tau^* \, ,
  \label{tauCP3}
  \end{align}
  where $n$ is an integer.
  The chain  $CP\rightarrow S\rightarrow CP^{-1}=\gamma'(S)$ imposes no further restrictions on $\tau_{CP}$.
  It is always possible to redefine the CP transformation in such a way that $n=0$
  by using the freedom of  $T$ transformation.
  Therefore,
 we  can define  the CP transformation of the modulus $\tau$ as
     \begin{align}
   \tau \xrightarrow{\, CP\, } -\tau^* \, .
   \label{tauCPfinal}
   \end{align}



\subsection{CP transformation of  modular multiplets}
Chiral superfields and modular forms   transform  
  in  Eqs.\,(\ref{modularforms}) and (\ref{chiralfields}),
 respectively, under a modular transformation.
 Chiral superfields also  transform
   in  Eq.\,(\ref{gCP})  under the  CP transformation.
  The  CP transformation of modular forms was given in Ref.\cite{Novichkov:2019sqv} as follows.
  Define a modular multiplet of the irreducible representation $\bf r$
  of $\Gamma_N$   with weight $k$ as $\bf Y^{\rm (k)}_{\bf r}(\tau)$,
  which is transformed as:
  \begin{align}
  \bf Y^{\rm (k)}_{\bf r}(\tau)
  \xrightarrow{\, {\rm CP} \, } Y^{\rm (k)}_{\bf r}(-\tau^*) \, ,
  \end{align} 
under the  CP transformation.
 The complex conjugated CP transformed modular forms
 $\bf Y^{\rm (k)*}_{\bf r}(-\tau^*)$ transform almost like the original multiplets
 $\bf Y^{\rm (k)}_{\bf r}(\tau)$  under a modular transformation, namely:

  \begin{align}
\bf Y^{\rm (k)*}_{\bf r}(-\tau^*) \xrightarrow{\ \gamma \ }
	Y^{\rm (k)*}_{\bf r}(-(\gamma\tau)^*) ={\rm (c\tau+d)^k} 
	\rho_{\bf {r}}^*({\rm u}(\gamma)) 
	Y^{\rm (k)*}_{\bf r}(-\tau^*) \, ,
\end{align}
where $u(\gamma)\equiv CP \gamma CP^{-1}$.
 Using the consistency condition of Eq.\,(\ref{consistency}), we obtain
   \begin{align}
 \bf X_r^T Y^{\rm (k)*}_{\bf r}(-\tau^*) \xrightarrow{\ \gamma \ }
 {\rm (c\tau+d)^k} \rho_{\bf {r}}(\gamma) 
 X_r^T Y^{\rm (k)*}_{\bf r}(-\tau^*) \, .
 \end{align}
 Therefore, if there exist a unique modular multiplet at 
  a level $N$, weight $k$ and representation $\bf r$,
  which is satisfied for $N=2$--$5$ with weight $2$,
  we can express the modular form $\bf Y^{\rm (k)}_{\bf r}(\tau)$ as:
  \begin{align}
\bf Y^{\rm (k)}_{\bf r}(\tau)= {\rm \kappa}  X_r^T Y^{\rm (k)*}_{\bf r}(-\tau^*) \, ,
\label{Yproportion}
\end{align}
where $\kappa$ is a  proportional coefficient.
Since $\bf Y^{\rm (k)}_{\bf r}(-(-\tau^*)^*)=\bf Y^{\rm (k)}_{\bf r}(\tau)$,
Eq.\,(\ref{Yproportion}) gives $\bf X_r^* X_r={\rm |\kappa|^2} \mathbb{1}_r $.
Therefore,
 the matrix $\bf X_r$ is a symmetric one, and $\kappa=e^{i \phi}$
 is a phase, which can be absorbed in the normalization of 
 modular forms.
 In conclusion, the CP transformation of modular forms  is given as:
 \begin{align}
\bf Y^{\rm (k)}_{\bf r}(\tau)\xrightarrow{\, {\rm CP} \, }
 Y^{\rm (k)}_{\bf r}(-\tau^*) =X_r  Y^{\rm (k)*}_{\bf r}(\tau)\, .
\end{align} 
It is also emphasized that $\bf X_r=\mathbb{1}_r$ satisfies the consistency
condition Eq.\,(\ref{consistency})
in a basis that  generators of $S$ and $T$ of $\Gamma_N$ are represented by symmetric matrices
because of 
$ \rho^*_{\bf {r}}(S)=  \rho^\dagger_{\bf {r}}(S)= \rho_{\bf {r}}(S^{-1})=
 \rho_{\bf {r}}(S)$ and 
 $ \rho^*_{\bf {r}}(T)=  \rho^\dagger_{\bf {r}}(T)= \rho_{\bf {r}}(T^{-1})$.

The CP transformations of  chiral superfields and modular multiplets
are summalized as follows:
  \begin{align}
 \tau \xrightarrow{\, {\rm CP} \, } -\tau^* \, , \qquad
 \psi (x)  \xrightarrow{\, {\rm CP} \, } X_r \overline \psi (x_P)\, , \qquad
 \bf Y^{\rm (k)}_{\bf r}(\tau)\xrightarrow{\, {\rm CP} \, } 
 Y^{\rm (k)}_{\bf r}(-\tau^*)  =X_r  Y^{\rm (k)*}_{\bf r}(\tau)\, ,
 \label{CPsummary}
 \end{align} 
 where  $\bf X_r=\mathbb{1}_r$ can be taken  in the basis of symmetric  generators of $S$ and $T$.
 We use this CP  transformation of modular forms to construct the CP invariant mass matrices in the next section.


\section{CP invariant lepton mass matrix in  $A_4$ modular symmetry}
\label{A4model}
In this section, we propose the CP invariant lepton mass matrix for introducing the $A_4$ modular symmetry.
The three generations of the left-handed lepton doublets are assigned to be an $A_4$ triplet $L$,
and the right-handed charged leptons $e^c$, $\mu ^c$, and $\tau ^c$ are $A_4$ singlets 
$\bf 1$, $\bf 1''$, and $\bf 1'$, respectively. The three generations of the right-handed neutrinos 
are also  assigned to be an $A_4$ triplet $N^c$. 
The weight of the superfields of left-handed leptons 
is fixed to be $-1$ as a standard.  The weight of right-handed neutrinos 
is also taken to be 
 $-1$  in order to give a Dirac neutrino mass matrix
 in terms of modular forms of weight $2$. 
On the other hand, weights of the right-handed charged leptons 
$e^c$, $\mu ^c$, and $\tau ^c$ are put $(k_e,\ k_\mu,\ k_\tau)$ in general. 
Weights of Higgs fields $H_u$, $H_d$ are  fixed to be $0$.
The representations and weights
 for MSSM fields and modular forms of weight $k$  are summarized in Table~\ref{tb:lepton}.

\begin{table}[h]
	\centering
	\begin{tabular}{|c||c|c|c|c|c|c|} \hline
		\rule[14pt]{0pt}{1pt}
		&$L$&$(e^c,\mu^c,\tau^c)$&$N^c$ &$H_u$&$H_d$&
		$ Y_{\bf 3}^{ (k)} $
		\\  \hline\hline 
		\rule[14pt]{0pt}{1pt}
		$SU(2)$&$\bf 2$&$\bf 1$& $\bf 1$ &$\bf 2$&$\bf 2$&$\bf 1$\\
		\rule[14pt]{0pt}{1pt}
		$A_4$&$\bf 3$& \bf (1,\ 1$''$,\ 1$'$)& $\bf 3$ &$\bf 1$&$\bf 1$&$\bf 3$\\
		\rule[14pt]{0pt}{1pt}
		weight & $ -1$ &$(k_e,\ k_\mu,\ k_\tau)$ & $-1$ & $0$ & $0$ &  $k$ \\ \hline
	\end{tabular}	
	\caption{ Representations and  weights
		 for MSSM fields and  relevant modular forms of weight $k$.
	}
	\label{tb:lepton}
\end{table}
Since we construct the CP invariant lepton  mass matrices with minimum number of parameters, 
we fix  weights
$k_e=-1,\ k_\mu=-3,\ k_\tau=-5$ for right-handed charged leptons.
Then,
we need modular forms of weight $2$, $4$ and $6$, $Y_{\bf 3}^{\rm (2)}$ 
$ Y_{\bf 3}^{\rm (4)}$ and $ Y_{\bf 3}^{\rm (6)}$.
For weight 4,  there are  five modular forms
two singlets and one triplet  of  $\rm A_4$.
Those are given in terms of weight $2$ modular forms
$Y_1(\tau)$, $Y_2(\tau)$ and $Y_3(\tau)$ as:
\begin{align}
&\begin{aligned}
{ Y^{\rm (4)}_{\bf 1}}(\tau)=Y_1(\tau)^2+2 Y_2(\tau) Y_3(\tau) \, , \qquad\quad\ \
{ Y^{\rm (4)}_{\bf 1'}}(\tau)=Y_3(\tau)^2+2 Y_1(\tau) Y_2(\tau) \, , 
\end{aligned}\nonumber \\
&\begin{aligned} 
{ Y^{\rm (4)}_{\bf 1''}}(\tau)=Y_2(\tau)^2+2 Y_1(\tau) Y_3(\tau)=0 \, , \qquad
{ Y^{\rm (4)}_{\bf 3}}(\tau)=
\begin{pmatrix}
Y_1^{(4)}(\tau)  \\
Y_2^{(4)}(\tau) \\
Y_3^{(4)}(\tau)
\end{pmatrix}
=
\begin{pmatrix}
Y_1(\tau)^2-Y_2(\tau) Y_3(\tau)  \\
Y_3(\tau)^2 -Y_1(\tau) Y_2(\tau) \\
Y_2(\tau)^2-Y_1(\tau) Y_3(\tau)
\end{pmatrix}\, . 
\end{aligned}
\label{weight4}
\end{align}
For the wight 6, we have  seven modular forms,
one singlet and two triplets  of  $\rm A_4$ as:
\begin{align}
&\begin{aligned}
 Y^{( 6)}_{\bf 1}=Y_1^3+ Y_2^3+Y_3^3 -3Y_1 Y_2 Y_3  \ , 
\end{aligned} \nonumber \\
\nonumber \\
&\begin{aligned}  Y^{(\rm 6)}_{\bf 3}\equiv 
\begin{pmatrix}
Y_1^{(6)}  \\
Y_2^{(6)} \\
Y_3^{(6)}
\end{pmatrix}
=(Y_1^2+2  Y_2 Y_3)
\begin{pmatrix}
Y_1   \\
Y_2 \\
Y_3
\end{pmatrix}\ , \qquad
\end{aligned}
\begin{aligned}  Y^{( 6)}_{\bf 3'}\equiv
\begin{pmatrix}
Y_1^{'(6)}  \\
Y_2^{'(6)} \\
Y_3^{'(6)}
\end{pmatrix}
=(Y_3^2+2 Y_1 Y_2)
\begin{pmatrix}
Y_3   \\
Y_1 \\
Y_2
\end{pmatrix}\ . 
\end{aligned}
\label{weight6}
\end{align}

Then, the $A_4$ invariant superpotential of the charged leptons, $w_E$,
by taking into account the modular weights is obtained as 
\begin{align}
w_E&=\alpha_e e^c H_d  Y^{ (2)}_{\bf 3}L+
\beta_e \mu^c H_d  Y^{ (4)}_{\bf 3}L+
\gamma_e \tau^c H_d  Y^{ (6)}_{\bf 3}L+
\gamma_e' \tau^c H_d  Y^{ (6)}_{\bf 3'}L~,
\label{chargedlepton}
\end{align}
where  $\alpha_e$,  $\beta_e$, $\gamma_e$, and $\gamma_e'$
 are constant parameters.
Under CP, the superfields transform as:
\begin{align}
e^c \xrightarrow{\,  CP\,}\, X_{\bf 1}^* \,\overline  e^c\, , \quad
\mu^c \xrightarrow{\,  CP\,} X_{\bf 1''}^*\, \overline\mu^c\, , \quad
\tau^c \xrightarrow{\,  CP\,}\, X_{\bf 1'}^* \,\overline \tau^c\, , \quad
L \xrightarrow{\,  CP\,}\, X_{\bf 3} \overline  L\, , \quad
H_d \xrightarrow{\,  CP\,}\,\eta_d\, \overline  H_d\, , 
\end{align} 
and we can take $\eta_d=1$ without loss of generality.
Since the representations of  
$S$ and $T$ are symmetric (see Appendic A), 
we can choose $X_{\bf 3}=\mathbb{1}$ 
and $X_{\bf 1}=X_{\bf 1'}=X_{\bf 1''}=\mathbb{1}$
as discused in Eq.\,(\ref{CPsummary}).

Taking $(e_L, \mu_L,\tau_L)$ in the flavor base,
the charged lepton mass matrix $M_E$  is simply written  as:    
\begin{align}
\begin{aligned}
M_E(\tau)=v_d \begin{pmatrix}
\alpha_e & 0 & 0 \\
0 &\beta_e & 0\\
0 & 0 &\gamma_e
\end{pmatrix}
\begin{pmatrix}
Y_1(\tau) & Y_3(\tau) & Y_2(\tau) \\
Y_2^{(4)}(\tau) & Y_1^{(4)}(\tau) & Y_3^{(4)}(\tau) \\
Y_3^{(6)}(\tau)+g_eY_3'^{(6)}(\tau) & Y_2^{(6)}(\tau)+g_eY_2'^{(6)}(\tau) & 
Y_1^{(6)}(\tau)+g_eY_1'^{(6)}(\tau)
\end{pmatrix}_{RL} \ ,
\end{aligned}
\label{ME(2)}
\end{align}
where $g_e=\gamma _e'/\gamma _e$, and $v_d$ is  VEV of the neutral component of $H_d$.
  The coefficients $\alpha_e$, $\beta_e$ and 
$\gamma_e$ are taken to be  real without loss of generality.
Under  CP transformation,  the mass matrix $M_E$ is transformed
following from  Eq.\,(\ref{ME(2)}) as:
\begin{align}
\begin{aligned}
M_E(\tau)&  \xrightarrow{\,  CP\,}  M_E (-\tau^*) = M_E^* (\tau)= \\
&
v_d \begin{pmatrix}
\alpha_e & 0 & 0 \\
0 &\beta_e & 0\\
0 & 0 &\gamma_e
\end{pmatrix}
\begin{pmatrix}
Y_1(\tau)^* & Y_3(\tau)^* & Y_2(\tau)^* \\
Y_2^{(4)}(\tau)^* & Y_1^{(4)}(\tau)^* & Y_3^{(4)}(\tau)^* \\
Y_3^{(6)}(\tau)^*+g_e^*Y_3'^{(6)}(\tau)^* & Y_2^{(6)}(\tau)^*+g_e^*Y_2'^{(6)}(\tau)^* & 
Y_1^{(6)}(\tau)^*+g_e^*Y_1'^{(6)}(\tau)^*
\end{pmatrix}_{RL} \ .
\end{aligned}
\label{CPME}
\end{align}

Let us discuss the  neutrino sector.
In Table~\ref{tb:lepton}, the $A_4$ invariant superpotential for the neutrino sector, $w_\nu$, is  given as:
\begin{align}
w_\nu &=w_D+w_N, \nonumber \\
w_D&=\gamma _\nu N^cH_u Y^{ (2)}_{\bf 3}L+
\gamma _\nu 'N^cH_u Y^{ (2)}_{\bf 3}L, \nonumber \\
w_N&=\Lambda N^cN^c Y^{ (2)}_{\bf 3}.
\label{eq:neutrino}
\end{align}
where $\gamma _\nu $ and  $\gamma _\nu '$ are Yukawa couplings,  and 
$\Lambda$ denotes a right-handed Majorana neutrino mass scale.
By putting $v_u$ for  VEV of the neutral component of $H_u$ 
and taking $(\nu_e, \nu_\mu,\nu_\tau)$ for neutrinos,
the Dirac neutrino mass matrix, $M_D$, is obtained as
\begin{align}
M_D=\gamma _\nu v_u\begin{pmatrix}
2Y_1 & (-1+g_D)Y_3 & (-1-g_D)Y_2 \\
(-1-g_D)Y_3 & 2Y_2 & (-1+g_D)Y_1 \\
(-1+g_D)Y_2 & (-1-g_D)Y_1 & 2Y_3\end{pmatrix}_{RL},
\label{MD}
\end{align}
where $g_D=\gamma _\nu '/\gamma _\nu $.
On the other hand the right-handed Majorana neutrino mass matrix, $M_N$ is written as follows:
\begin{align}
M_N=\Lambda\begin{pmatrix}
2Y_1 & -Y_3 & -Y_2 \\
-Y_3 & 2Y_2 & -Y_1 \\
-Y_2 & -Y_1 & 2Y_3\end{pmatrix}_{RR}.
\label{MR}
\end{align}
By using the type-I seesaw mechanism, the effective neutrino mass matrix, $M_\nu$ is obtained as
\begin{align}
M_\nu=M_D^{\rm T}M_N^{-1}M_D ~.
\label{seesaw}
\end{align}

In a CP conserving modular invariant theory, both CP and modular symmetries are broken spontaneously by  VEV of the modulus $\tau$.
However, there exist certain values of $\tau$ which conserve CP while breaking the modular symmetry.
Obviously, this is the case if
$\tau$ is left invariant by CP, i.e.
\begin{align}
\tau \xrightarrow{\,  CP\,}   -\tau^*=\tau\, \, ,
\label{CPtau}
\end{align}
which indicates $\tau$ lies on the imaginary axis, ${\rm Re} [\tau]=0$.
In addition to ${\rm Re} [\tau]=0$, 
CP is conserved at the boundary of the fundamental domain.

Due to Eq.\,(\ref{CPsummary}),
 one then has
\begin{align}
M_E(\tau)=M_E^*(\tau)\, ,\qquad\qquad M_\nu(\tau)=M_\nu^*(\tau) \, ,
\label{CPMassmatrix}
\end{align}
 if   $g_{e}$ and  $g_{D}$ are taken to be   real.
Therefore,
 the source of the CP violation is only non-trivial ${\rm Re}[\tau]$
after breaking the modular symmetry.
In the next section, we present a numerical analysis of the CP violation by
 fixing  the modulus $\tau$ with   real $g_{e}$ and  $g_{D}$.

\section{Numerical results of leptonic CP violation}

 We have presented the CP invariant lepton mass matrices
  in the $A_4$ modular symmetry.
  The tiny neutrino masses are given via type-I seesaw.
  The CP symmetry is broken spontaneously by  VEV of the modulus $\tau$.
  Thus,  VEV of $\tau$ breaks the CP invariance as well as the modular invariance.
  The source of the CP violation is  the real part of $\tau$.
  Indeed, the spontaneous CP violation
  is realized  by fixing  $\tau$.
  Then, the  Dirac CP phase and Majorana phases
   are predicted clearly  with reproducing observed lepton mixing angles and two neutrino mass squared differences.
 The predictive  CP  phases are possibly  linked to the phase 
 of  the leptogenesis \cite{Fukugita:1986hr}.


   Our parameters are real ones 
      $\alpha_e$, $\beta_e$, $\gamma_e$, $\gamma'_e$, $\gamma_\nu$, $\gamma'_\nu$ and  $\Lambda$
       in addition to the complex $\tau$. 
       Observed input data are three charged lepton masses,
        three flavor mixing angles and two neutrino mass squared differences. 
        Since  $\gamma_\nu$ and $\Lambda$ appear only with  
          the combination $\gamma_\nu^2/\Lambda$ in the neutrino mass matrix,
           the input data determine completely
       our parameters apart from error-bars  of the experimental data.
       Therefore, the lepton mixing angles, the Dirac phase  and Majorana phases are predicted in the restricted ranges.

 As the input charged lepton masses, 
we take Yukawa couplings of charged leptons 
at the GUT scale $2\times 10^{16}$ GeV,  where $\tan\beta=5$ is taken as a bench mark
\cite{Antusch:2013jca, Bjorkeroth:2015ora}:
\begin{eqnarray}
y_e=(1.97\pm 0.024) \times 10^{-6}, \quad 
y_\mu=(4.16\pm 0.050) \times 10^{-4}, \quad 
y_\tau=(7.07\pm 0.073) \times 10^{-3},
\end{eqnarray}
where lepton masses are  given by $m_\ell=y_\ell v_H$ with $v_H=174$ GeV.

\begin{table}[H]
	\begin{center}
		\begin{tabular}{|c|c|c|}
			\hline
			      \rule[14pt]{0pt}{2pt}
\ observable\       &                 best fit\,$\pm 1\,\sigma$ for NH                  &           best fit\,$\pm 1\,\sigma$  for IH            \\ \hline
			          \rule[14pt]{0pt}{2pt}				                                          $\sin^2\theta_{12}$               &                     $0.304^{+0.012}_{-0.012}$                     &               $0.304^{+0.013}_{-0.012}$                \\                 \rule[14pt]{0pt}{2pt}
			$\sin^2\theta_{23}$   &                     $0.573^{+0.016}_{-0.020}$                     &               $0.575^{+0.016}_{-0.019}$                \\
			          \rule[14pt]{0pt}{2pt}
			              $\sin^2\theta_{13}$               &                  $0.02219^{+0.00062}_{-0.00063}$            
			              &    $0.02238^{+0.00063}_{-0.00062}$    \\ 
			          \rule[14pt]{0pt}{2pt}
$\Delta m_{\rm sol }^2$  & $7.42^{+0.21}_{-0.20}\times 10^{-5}{\rm eV}^2$  &     $7.42^{+0.21}_{-0.20}\times 10^{-5}{\rm eV}^2$     \\ 
\rule[14pt]{0pt}{2pt}
$\Delta m_{\rm atm}^2$ & \ \ \ \ $2.517^{+0.026}_{-0.028}\times 10^{-3}{\rm eV}^2$ \ \ \ \ & $-2.498^{+0.028}_{-0.028}\times 10^{-3}{\rm eV}^2$ \ \ \\  \hline
		\end{tabular}
\caption{The best fit\,$\pm 1\,\sigma$ of neutrino  parameters from NuFIT 5.0
			for NH and IH 
		\cite{Esteban:2020cvm}.
		}
		\label{DataNufit}
	\end{center}
\end{table}
\vskip -0.5 cm

We also input  the   lepton mixing angles and neutrino mass parameters
which are given by NuFit 5.0 in Table \ref{DataNufit} \cite{Esteban:2020cvm}.
In our analysis, the Dirac CP phase $\delta_{CP}$
(see Appendix C) is output because its observed range
 is too wide at $3\,\sigma$ confidence level.
We investigate  two possible cases of neutrino masses $m_i$, which are
the normal  hierarchy (NH), $m_3>m_2>m_1$, and the  inverted  hierarchy (IH),
$m_2>m_1>m_3$.
Neutrino masses and  
the PMNS matrix $U_{\rm PMNS}$ \cite{Maki:1962mu,Pontecorvo:1967fh} 
are obtained by diagonalizing 
$M_E^\dagger M_E$ and $M_\nu^\dagger M_\nu$.
We also investigate the effective mass for the $0\nu\beta\beta$ decay,
$\langle m_{ee} \rangle$ (see Appendix C)
and 
the sum of three neutrino  masses  $\sum m_i$  since
it is constrained by the recent cosmological data,
which is  the upper-bound $\sum m_i\leq 120$\,meV obtained at the 95\% confidence level
\cite{Vagnozzi:2017ovm,Aghanim:2018eyx}.

  Let us  discuss numerical results for NH of neutrino masses.
We scan $\tau$ in the fundamental domain of $SL(2,Z)$.
The real parameters  $g_e$ and $g_D$   are scanned in $ [-10,\,10]$.
    As a measure of good-fit, we adopt the sum of one-dimensional 
$\chi^2$ functions for five accurately known  observables
$\Delta m_{\rm atm}^2$, $\Delta m_{\rm sol}^2$,
$\sin^2\theta_{12}$, $\sin^2\theta_{23}$ and $\sin^2\theta_{13}$
 in NuFit 5.0 \cite{Esteban:2020cvm}. In addition, we employ Gaussian approximations for fitting
 $m_e$,  $m_\mu$ and $m_\tau$.
 
\begin{figure}[H]
  	\begin{minipage}[]{0.47\linewidth}
  		\vspace{5mm}
  		\includegraphics[{width=\linewidth}]{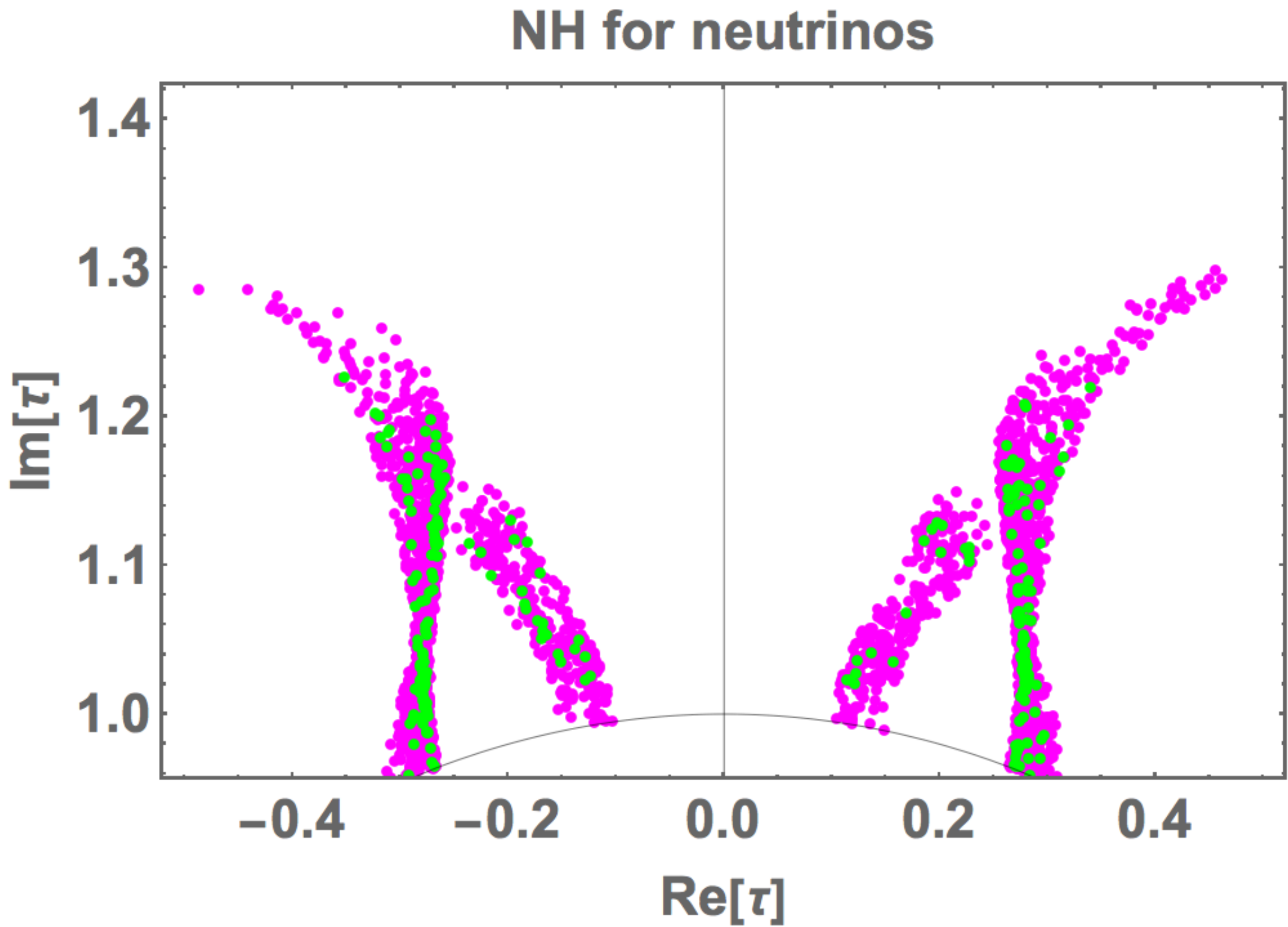}
  		\caption{
  			Allowed regions of $\tau$ for  NH, where
  green  and  magenta points  correspond to $\sqrt{\chi^2}\leq 2$ 
  and $ 3$,
  			  respectively.
  			The  solid curve is the boundary of the fundamental domain, $|\tau|=1$.}
  			\label{tau}
  	\end{minipage}
  	\hspace{5mm}
  		\begin{minipage}[]{0.47\linewidth}
  			\vspace{2mm}
  			\includegraphics[{width=\linewidth}]{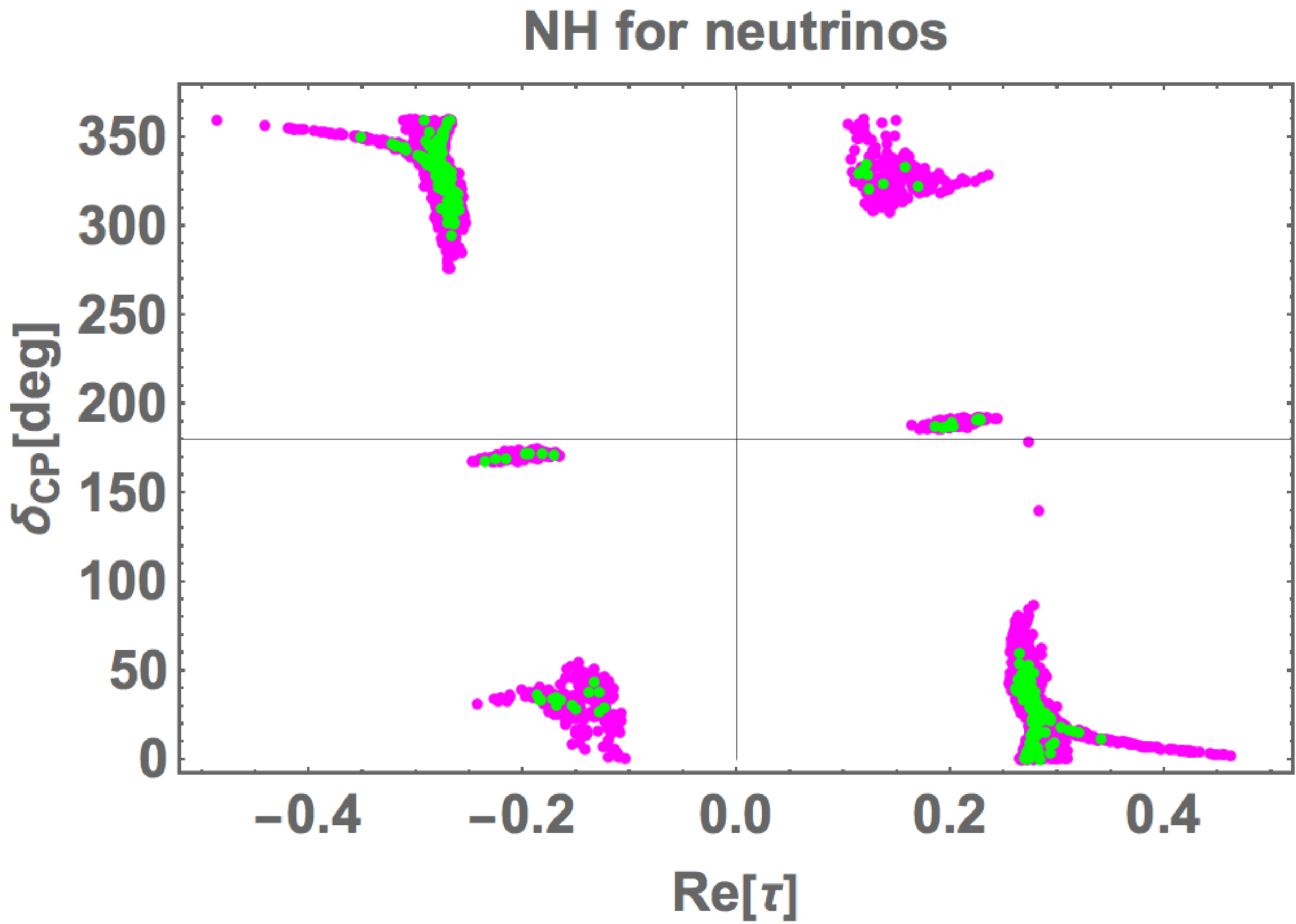}
  			\caption{
  				Prediction of Dirac phase $\delta_{CP}$ versus
  				${\rm Re}\, [\tau]$ for NH.
  				There are six regions, which are almost  symmetric  with respect to the point (${\rm Re}\, [\tau]=0,\, \delta_{CP}=180^\circ$).
  				Colors denote same ones in Fig.\,\ref{tau}.}
  			\label{delcp-reltau}
  		\end{minipage}
  \end{figure}

   In Fig.\,\ref{tau},  we show the allowed region  on the 
  ${\rm Re}\, [\tau]$\,--\,${\rm Im}\,[\tau]$ plane, where three mixing angles,
  $\Delta m_{\rm atm}^2$, $\Delta m_{\rm sol}^2$
   and charged lepton masses are consistent with observed ones.
  The green and magenta regions correspond to  $\sqrt{\chi^2}\leq 2$ and  
  $3$ 
  , respectively.
  The predicted range of  $\tau$ is
  in ${\rm Re}\,[\tau]=\pm [0.1,0.5]$  and
  ${\rm Im}\,[\tau]=[0.96,1.30]$ at   $\sqrt{\chi^2}\leq 3$
   (magenta).
 
   Due to the rather broad range of  ${\rm Re}\,[\tau]$, the 
   predictive Dirac    CP phase $\delta_{CP}$,
 which is defined in Appendix C,  is not so restricted.
 In Fig.\,\ref{delcp-reltau}, we show a prediction of $\delta_{CP}$ versus ${\rm Re}\,[\tau]$.
 It is remarked that  $\delta_{CP}$ is predicted in  six 
 regions depending on the sign of ${\rm Re}\,[\tau]$.
  Those are
  $[0^\circ,50^\circ]$, $[170^\circ,175^\circ]$,
   $[280^\circ,360^\circ]$ for ${\rm Re}\,[\tau]<0$,
    and 
    $[0^\circ,80^\circ]$, $[185^\circ,190^\circ]$,
    $[310^\circ,360^\circ]$ for ${\rm Re}\,[\tau]>0$
    at $\sqrt{\chi^2}\leq 3$
   (magenta).
  These are almost  symmetric  with respect to the point 
  (${\rm Re}\, [\tau]=0,\, \delta_{CP}=180^\circ$).
  This prediction is consistent with the result of  global fit of  
  NuFit 5.0  \cite{Esteban:2020cvm}:
  \begin{eqnarray}
  \delta_{CP}=197^{\circ \, +27^\circ}_{\ \, -24^\circ} \, .
  \end{eqnarray}

    \begin{figure}[H]
 \begin{minipage}[]{0.47\linewidth}
 	\includegraphics[{width=\linewidth}]{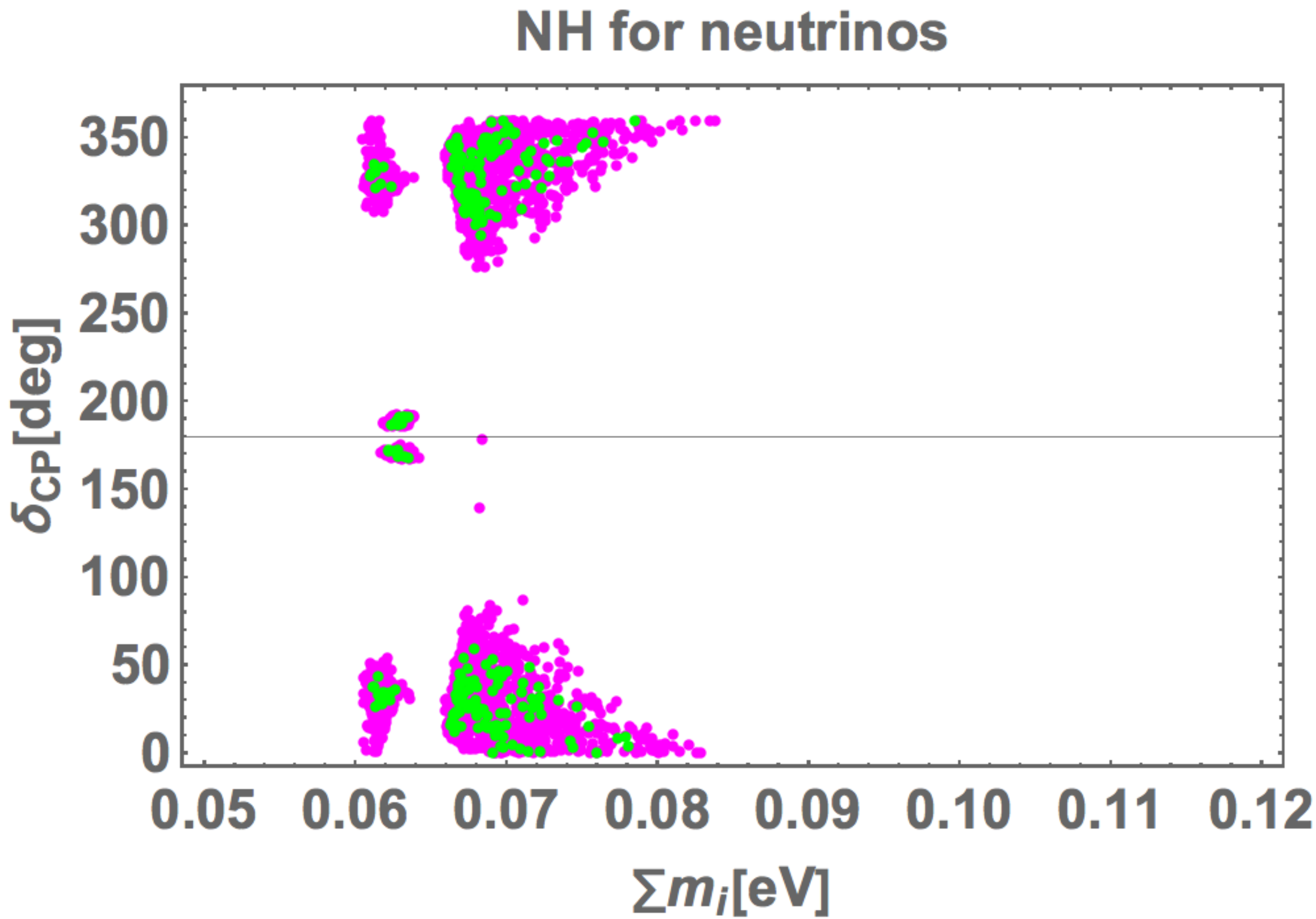}
 	\caption{Predicted six regions  of Dirac phase $\delta_{CP}$ versus the sum of neutrino masses  $\sum m_i$ for NH.
 		Colors denote same ones in Fig.\,1.}
 	\label{Dirac1}
 \end{minipage}
  	\hspace{5mm}
  	\begin{minipage}[]{0.47\linewidth}
  		\includegraphics[{width=\linewidth}]{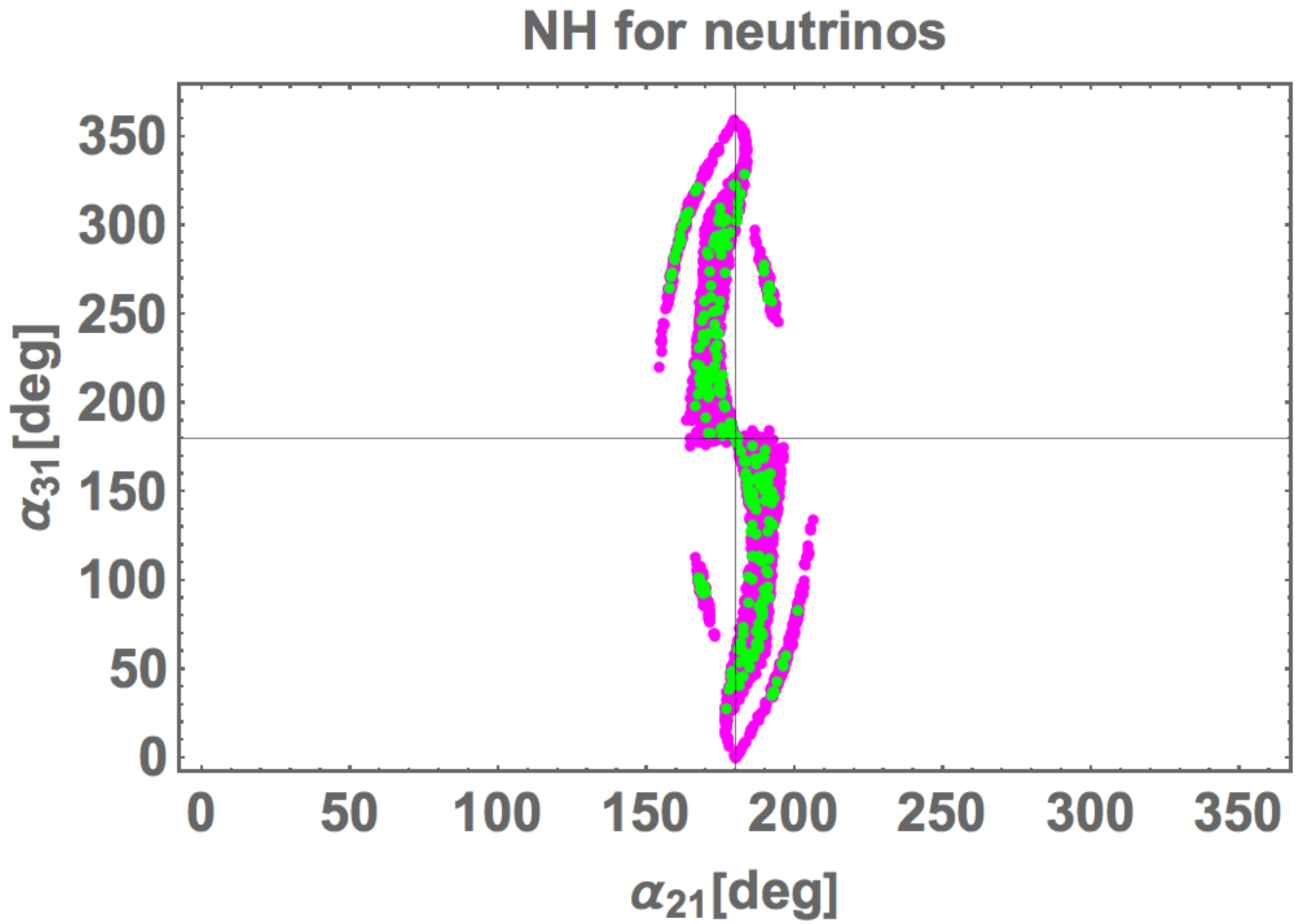}
  		\caption{ Predicted Majorana phases $\alpha_{21}$
  			and $\alpha_{31}$ for NH.
  			Colors denote same ones in Fig.\,1.
  		}
  	\label{Majorana}
  	\end{minipage}
  \end{figure}
  In Fig.\,\ref{Dirac1},
  we show a prediction of $\delta_{CP}$ versus $\sum m_i$.
  It is also found different six predicted regions.
  The sum of neutrino masses $\sum m_i$ is restricted 
  in the   narrow range $[60,\,84]$\,meV at $\sqrt{\chi^2}\leq 3$
    (magenta).
  It  is consistent with 
  the cosmological bound $120$\, meV in 
  the minimal cosmological model, ${\rm \Lambda CDM}+\sum m_i$.
  \cite{Vagnozzi:2017ovm,Aghanim:2018eyx}. 

 In Fig.\,\ref{Majorana}, we show the prediction of Majorana phases $\alpha_{21}$ and $\alpha_{31}$, which are defined by Appendix C. 
 The predicted $\alpha_{21}$ is around $180^\circ$,
 but $\alpha_{31}$ is distributed in the full range
 of  $[0^\circ,360^\circ]$ at $\sqrt{\chi^2} \leq 3$
  (magenta).

 
 We can calculate   the effective mass 
 $\langle m_{ee}\rangle$ for the $0\nu\beta\beta$ decay
  by using  the Dirac CP phase and Majorana phases as seen in Appendix C.
 The predicted $\langle m_{ee}\rangle$ is
  \begin{eqnarray}
 \langle m_{ee}\rangle = [0.003,\,3]\,{\rm meV} \, ,
 \end{eqnarray}
 at  $\sqrt{\chi^2} \leq 3$.
 It is difficult to reach this value in  the future experiments  of the neutrinoless double beta decay.

  We have checked the correlation between three mixing angles
  and   $\delta_{CP}$  plane.
  The predicted $\delta_{CP}$ is correlated  weakly with 
  $\sin^2\theta_{23}$ in our model.
  It has the broadest ranges of
  $[0^\circ,80^\circ]$ and  $[280^\circ,360^\circ]$
   at the best fit value of $\sin^2\theta_{23}=0.573$,
   but ranges of  $[0^\circ,30^\circ]$ and  $[330^\circ,360^\circ]$ are excluded  near  the observed upper bound of $\sin^2\theta_{23}$.
   On the other hand, there are no correlations
    among $\sin^2\theta_{12}$, $\sin^2\theta_{13}$ and $\delta_{CP}$.

 We show  the best fit sample for NH
in Table \ref{samplelepton}, where numerical values of 
parameters and output are listed. As a measure of goodness of fit, 
 we show  square root of the sum of one-dimensional $\chi^2$ functions.


 We have also scanned the parameter space for the case of IH of neutrino masses.
 We have found parameter sets which 
 reproduce the observed masses and  three mixing angles $\sin^2\theta_{23}$, $\sin^2\theta_{12}$, and $\sin^2\theta_{13}$ at
  $\sqrt{\chi^2} \leq 5$.
 However, there is no parameter sets below 
    $\sqrt{\chi^2}=4$.

\begin{figure}[H]
	\begin{minipage}[]{0.47\linewidth}
		\vspace{5mm}
		\includegraphics[{width=\linewidth}]{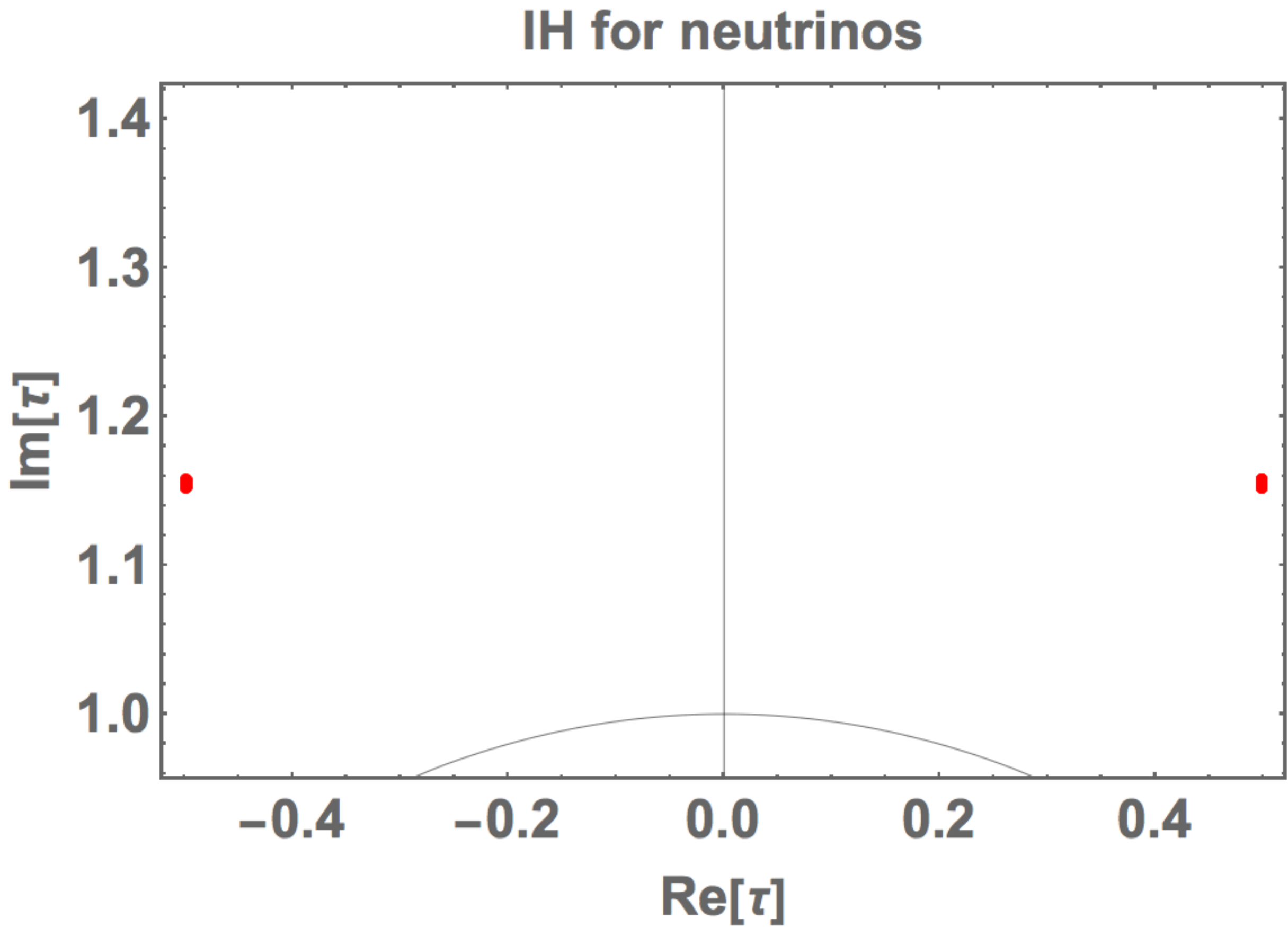}
		\caption{Allowed regions of $\tau$ for  IH 
			at    $\sqrt{\chi^2}=4$--$5$.
		}
	\label{tauIH}
	\end{minipage}
	\hspace{5mm}
	\begin{minipage}[]{0.47\linewidth}
		\vspace{2mm}
		\includegraphics[{width=\linewidth}]{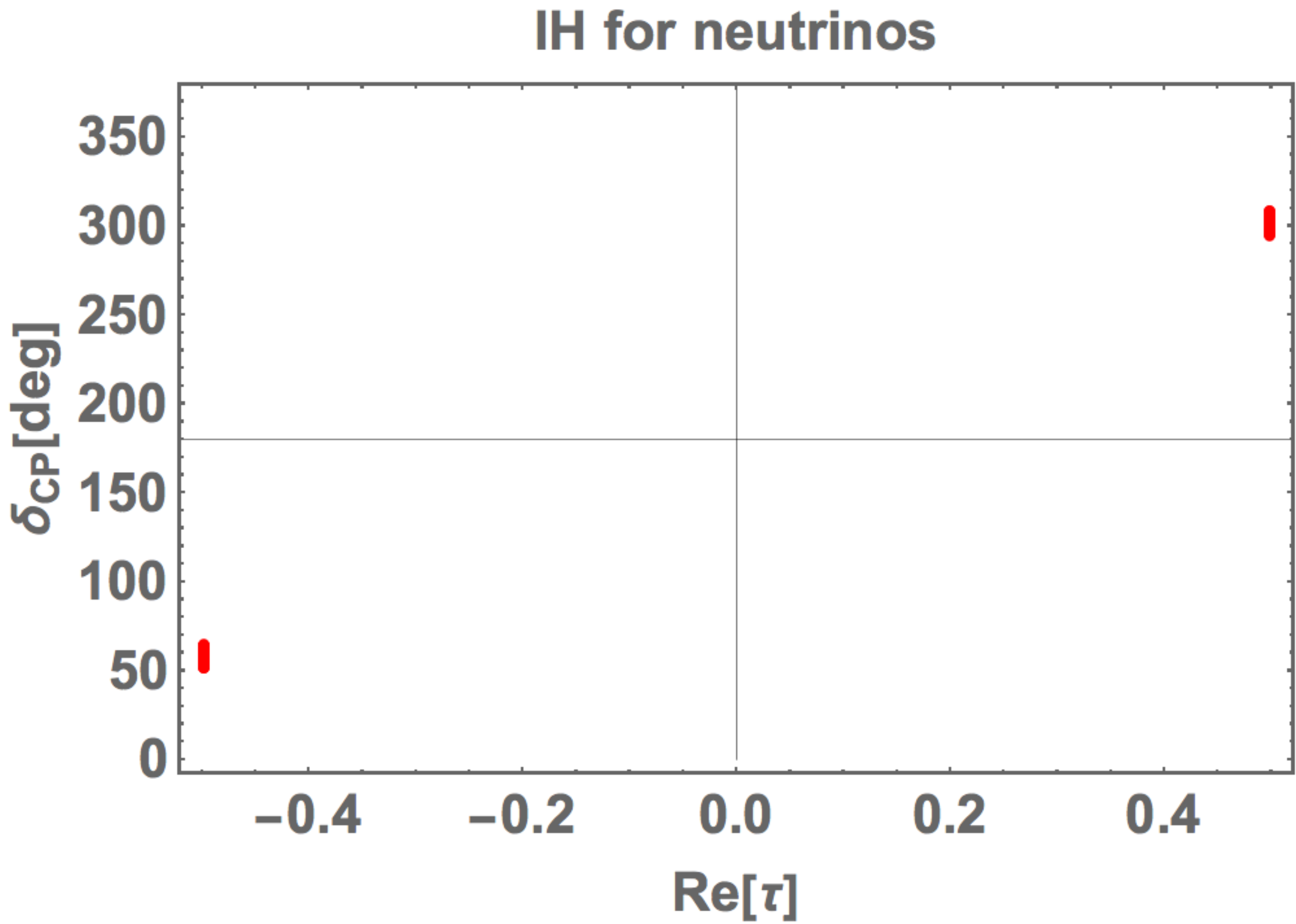}
		\caption{Prediction of Dirac phase $\delta_{CP}$
			versus ${\rm Re}\,[\tau]$ for  IH at $\sqrt{\chi^2}=4$--$5$
		}
		\label{deltaCPIH}
			\end{minipage}
\end{figure}

 The allowed region of $\tau$ is restricted in the narrow regions.
 As shown in Fig.\,\ref{tauIH}, the predicted range of   
 ${\rm Im}\,[\tau]$ is $[1.15,1.16]$ at    $\sqrt{\chi^2}=4$--$5$ 
 and ${\rm Re}\,[\tau]$ is close to $\pm 0.5$.
\begin{table}[H]
	\centering
	\begin{tabular}{|c|c|c|} \hline 
			\rule[14pt]{0pt}{0pt}
		&  NH & IH\\  \hline
			\rule[14pt]{0pt}{0pt}
		$\tau$&   $  -0.2637 + 1.1549  \, i$ & $0.4984 + 1.1553 \, i$ \\ 
		\rule[14pt]{0pt}{0pt}
		$g_D$ &$ -1.29$ & $1.74$\\
		\rule[14pt]{0pt}{0pt}
		$g_e$  &  $-1.01$& $1.68\times 10^{-7}$\\
		\rule[14pt]{0pt}{0pt}
		$\beta_e/\alpha_e$ & $4.66\times 10^{-2}$ &$3.64\times 10^{-2}$ \\
		\rule[14pt]{0pt}{0pt} 
		$\gamma_e/\alpha_e$ &  $ 11.9$ &$7.35\times 10^{-4}$ \\
		\rule[14pt]{0pt}{0pt}
		$\sin^2\theta_{12}$ & $ 0.305$& $0.309$\\
		\rule[14pt]{0pt}{0pt}
		$\sin^2\theta_{23}$ &  $ 0.571$& $0.494$\\
		\rule[14pt]{0pt}{0pt}
		$\sin^2\theta_{13}$ &  $ 0.0220$&$0.0222$	\\
		\rule[14pt]{0pt}{0pt}
		$\delta_{CP}$ &  $317^\circ$& $300^\circ$\\
		\rule[14pt]{0pt}{0pt}
		$[\alpha_{21},\,\alpha_{31}]$ &  $[189^\circ,\,64^\circ]$& 
		 $[116^\circ,\,270^\circ]$	\\	
		\rule[14pt]{0pt}{0pt}
		$\sum m_i$ &  $67.3$\,meV &	 $145$\,meV\\
		\rule[14pt]{0pt}{0pt}
		$\langle m_{ee} \rangle$ &  $0.18$\,meV& $35.5$\,meV \\
		\rule[14pt]{0pt}{0pt}
		$\sqrt{\chi^2}$ &  $1.39$ & $4.27$ \\
		\hline
	\end{tabular}
	\caption{Numerical values of parameters and observables
		at the best fit points of NH and IH.}
	\label{samplelepton}
\end{table}


In Fig.\,\ref{deltaCPIH}, we show the allowed region on 
the ${\rm Re}\,[\tau]$\,--\,$\delta_{CP}$  plane.
The predicted  $\delta_{CP}$ is severely restricted as in  $[50^\circ,70^\circ]$ and $[290^\circ,310^\circ]$.
 This prediction is consistent with the result of  global fit of 
 IH in NuFit 5.0, 
 $\delta_{CP}=282^{\circ \, +26^\circ}_{\ \, -30^\circ}$.
In addition,
$\sin^2\theta_{23}$ is restricted  in  $[0.505,0.515]$
\cite{Esteban:2020cvm}.

On the other hand,
the sum of neutrino masses $\sum m_i$ is restricted 
in the   narrow range $[143,\,147]$\,meV  at    $\sqrt{\chi^2}=4$--$5$.
Therefore, the case of IH will be  excluded by the 
improved  cosmological bound in the near future.
The predicted $\langle m_{ee}\rangle$ is given as
$\langle m_{ee}\rangle = [35,\,36]\,{\rm meV}$.

We also present the best fit  set of IH  in Table \ref{samplelepton},
where values of relevant parameters are listed compared with the values
of the NH case.

In our numerical calculations, we have not included  the RGE effects
 in the lepton mixing angles.
 We suppose that those corrections  are very small between 
 the electroweak  and GUT scales.
This assumption is  justified  well in the case of $\tan\beta\leq 10$
unless neutrino masses are almost degenerate \cite{Criado:2018thu}.

 
We have presented  CP invariant lepton  mass matrices with minimum number of parameters by fixing  weights
($k_e=-1,\ k_\mu=-3,\ k_\tau=-5$) for right-handed charged leptons.
Therefore,
the charged lepton mass matrix is given
 by   modular forms of weight $2$, $4$ and $6$, $Y_{\bf 3}^{ (2)}$,
 $ Y_{\bf 3}^{ (4)}$ and $ Y_{\bf 3}^{ (6)}$.
However, this choice for weights
 is not a unique one even if we consider the seesaw model
 with  minimum number of parameters. 
  We have examined alternative three choices:
  ($k_e=-1,\ k_\mu=-1,\ k_\tau=-5$),
  ($k_e=-1,\ k_\mu=-1,\ k_\tau=-7$) and
   ($k_e=-1,\ k_\mu=-3,\ k_\tau=-7$).
 The correponding  charged lepton mass matrices are presented in Appendix D.
 In those three models,  we have also  obtained successful numerical results,
 which are not so different  from above ones.
 We present  samples of parameter sets for each model in Appendix D
 for NH of neutrino masses. The cases of IH are  omitted
 since  their $\sqrt{\chi^2}$ are larger than $4$.
 Our study of the leptogenesis is focused on the case of  weights
 ($k_e=-1,\ k_\mu=-3,\ k_\tau=-5$) in the next section.

\section{Leptogenesis}
The BAU  at the present universe is measured very precisely by the cosmic microwave background radiation as\cite{Aghanim:2018eyx}:
\begin{equation}
Y_B =\frac{n_B}{s}= (0.852-0.888)\times 10^{-10} \,,
\label{observedYB}
\end{equation}
at $3\,\sigma$ confidence level,
where  $Y_B$ is defined by the ratio between the number density of baryon asymmetry $n_B$ and the entropy density $s$. 
 One of the most attractive scenarios for
baryogenesis is the canonical leptogenesis scenario \cite{Fukugita:1986hr} in which the decays of right-handed neutrinos can
generate the lepton asymmetry that is partially converted into the baryon asymmetry via the sphaleron process \cite{Kuzmin:1985mm}. The sign and magnitude of the BAU are predicted by the masses and Yukawa coupling constants of right-handed neutrinos. 
If their masses are hierarchical, the lightest one must be ${\cal O}(10^9)$ GeV \cite{Davidson:2002qv}
to explain the BAU. 
The sign of the BAU depends on the CP phase strucuture  in the lepton
mass matrices. In general, the sign of the BAU cannot be predicted uniquely even if the  Dirac and Majorana CP phases are determined. This is because there exist generally one or more additional phases associated with right-handed neutrinos which  decouple from the low energy phenomena even if right-handed neutrinos are sufficiently heavy. However, 
 our predictive   Dirac and Majorana CP phases are  linked  to the BAU because the CP violation  is originated from only  $\tau$ in our  model with  $A_4$ modular symmetry.

Let us discuss the leptogenesis by decays of right-handed neutrinos in our model.  
Since  the mass ratios of right-handed neutrinos are not so large,  and then 
we have to include the effects of all three right-handed neutrinos to the leptogenesis.
For simplicity, we assume that the reheating temperature of inflation
is sufficiently higher than the mass of the heaviest right-handed neutrino
and that the initial abundances of all right-handed neutrinos are zero. 
On the other hand, the mass degeneracy of the right-handed neutrinos is not so large,
and so the resonant enhancement of the leptogenesis \cite{Pilaftsis:2003gt,Asaka:2018hyk} does not occur.
Thus, we shall use the formalism based on the Boltzmann equations to estimate the asymmetries.
Moreover, as we show below, the required masses of right-handed 
neutrinos are ${\cal O}(10^{13})$~GeV, and so we can apply the simple one-flavor approximation
of the leptogenesis. Therefore, we neglect the so-called flavor effect~\cite{Abada:2006fw,Nardi:2006fx,Abada:2006ea,Blanchet:2006be,Pascoli:2006ie,Pascoli:2006ci,Moffat:2018smo,DeSimone:2006nrs}. Furthermore, to achieve successful leptogenesis via the decay of such heavy right-handed neutrinos, the reheating temperature must be higher than $\mathcal{O}(10^{13})$~GeV. In the framework of supersymmetry~(SUSY), such high reheating temperature can cause the overproduction of gravitinos, which is called the gravitino problem \cite{Khlopov:1984pf, Ellis:1984eq}. However, in our scenario, we assume that SUSY is broken at close to the Planck scale. In this situation, SUSY particles, including the gravitino, have masses around the Planck scale. Therefore, gravitino cannot be thermally produced after inflation. So the constraint on the reheating temperature due to the gravitino problem can be eliminated.

 The flavor structure of our model appears in the 
 $\mathrm {CP}$ asymmetry parameter  $\varepsilon_I$,
  which is:
\begin{align}
\varepsilon_{I}
=\frac{\Gamma\left(N_{I} \rightarrow L+\overline{H_u}\right)-\Gamma\left(N_{I} \rightarrow \overline{L}+H_u\right)}{\Gamma\left(N_{I} \rightarrow L+\overline{H_u}\right)+ \Gamma\left(N_{I} \rightarrow \overline{L}+H_u\right) 
}\  .
\label{asymmetric-parameter}
\end{align}
It is proportional to the imaginary part of Yukawa couplings as: 
\begin{equation}
\varepsilon_{I}\propto  \sum_{J \neq I}
{\rm Im}\{(y_\nu y_\nu^\dagger)_{IJ}\}^2 \,  .
\label{asym0}
\end{equation}
Here, $ y_\nu y_\nu^\dagger$ is given  by   the Dirac neutrino mass matrix $M_D$
in the real diagonal base of the right-handed Majorana neutrino  mass matrix $M_N$ as follows:
\begin{align}
 y_\nu y_\nu^\dagger=\frac{1}{v_u^2}V_R^\dagger\, (M_D M_D^\dagger)\, V_R \, , \qquad {\rm with } \quad
  V_R^\dagger \,(M_N M_N^\dagger)\, V_R={\rm diag}\,(M_1^2, M_2^2, M_3^2) \, ,
\end{align}
where $M_D$ and $M_N$ are  given in Eqs.\,(\ref{MD}) 
and (\ref{MR}), respectively, and
 $M_1$, $M_2$ and $M_3$ are real.


The Boltzmann equations are then solved numerically and the total lepton asymmetry 
$Y_L$ from the decays of right-handed neutrinos is estimated.  
The present baryon asymmetry can be estimated 
as $Y_B= -8/23 Y_L$ for the two Higgs doublets (see Appendix E).

In our model, the phases in the PMNS matrix and the high energy phases associated with 
right-handed neutrinos are originated in the modulus
$\tau$.  In this situation, there may exist the correlations between 
the phases in the PMNS matrix and the yield of the BAU.

Since the best-fit point of the modulus $\tau$
	is rather close to the fixed point  $\tau=i$
	for NH as seen in Table \ref{samplelepton},	we can  calculate approximately  the asymmetry parameter $\varepsilon_{I}$ of Eq.(\ref{asym0})
	in terms of a small complex  parameter $\epsilon$,
	which is defined as  $\tau=i+\epsilon$ in perturbation.
	This analytic calculation is possible
	due to the simple Dirac neutrino mass matrix at  $\tau=i$.
	It is found that 
	the leading term of $ {\rm Im}\{(y_\nu y_\nu^\dagger)_{Ij}\}^2$
	is given by $ {\rm Im}\,[\epsilon^2]$.
On the other hand, 
 it is almost impossible to give an approximate
  form of the CP phase $\delta_{CP} $ or 
   the CP violating measure $J_{CP}$ at  low energy.
  Since the right-handed Majorana neutrino mass matrix $M_N$
  gives one massless and two degenerated masses  at $\tau=i$ 
  in our model, its inverse  is a singular one at $\tau=i$.
   After seesaw, the left-handed Majorana mass matrix
  is unstable at nearby $\tau=i$ for perturbation.
  Indeed, we could not  obtain  a reliable analytic expression
  for $J_{CP}$.
  Therefore, we study the correlations between 
   the CP phase $\delta_{CP}$ and the yield of the BAU
    in numerical calculations.
\section{Baryon asymmetry}\label{sec:yieldBAU}
\begin{figure}[H]
 	\begin{minipage}[]{0.47\linewidth}
 		\vspace{5mm}
 		\includegraphics[{width=\linewidth}]{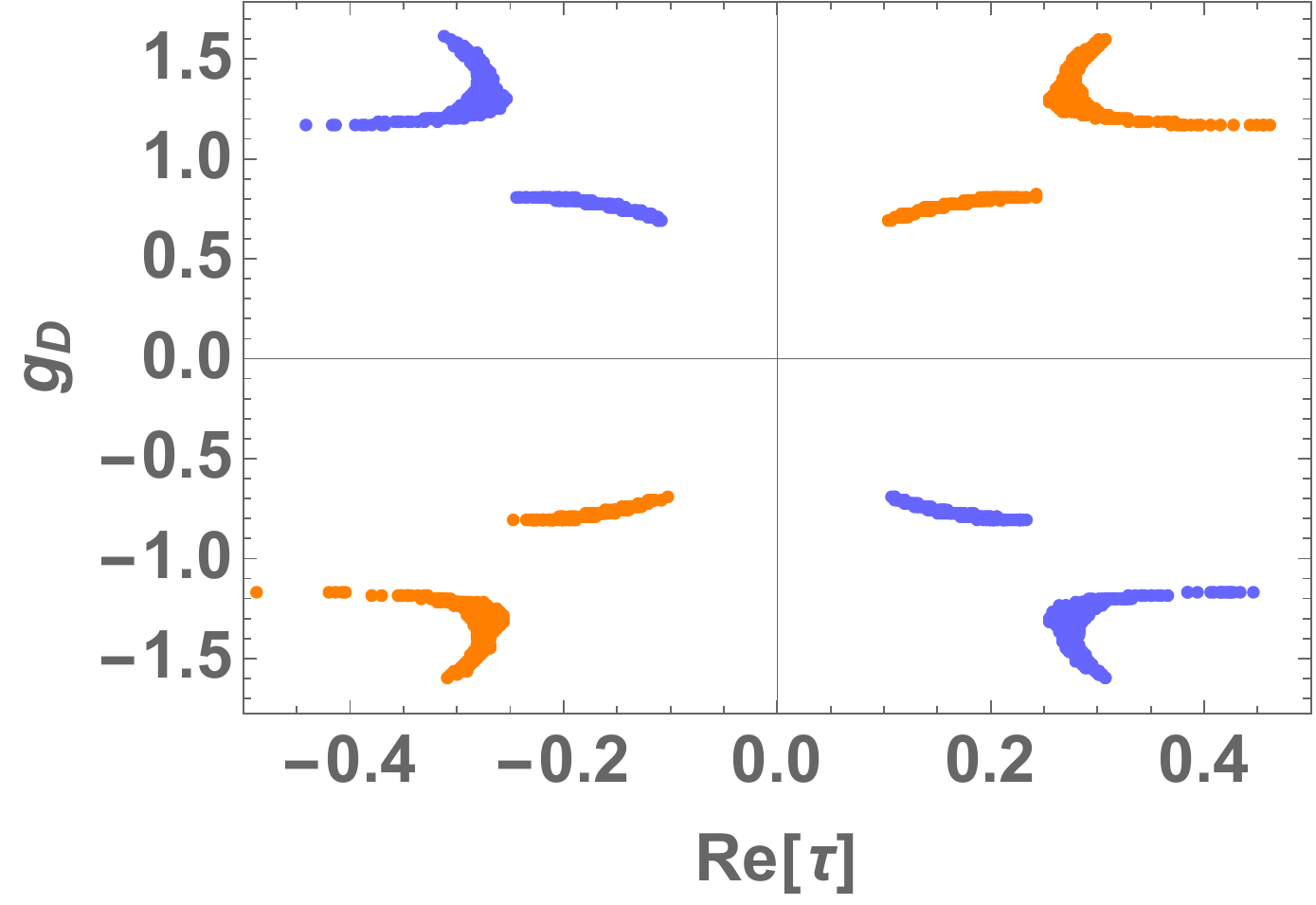}
 		\caption{The sign of $Y_B$ and regions of
 			${\rm Re}[\tau]$--$g_D$.	Orange and blue points
 			denote positive and negative $Y_B$, respectively.
 			Points  correspond to the output
 			of section 4 at    $\sqrt{\chi^2}\leq 3$
 			}
 		\label{gDRetau}
 	\end{minipage}
 	\hspace{5mm}
 	\begin{minipage}[]{0.47\linewidth}
 		\vspace{5mm}
 		\includegraphics[{width=\linewidth}]{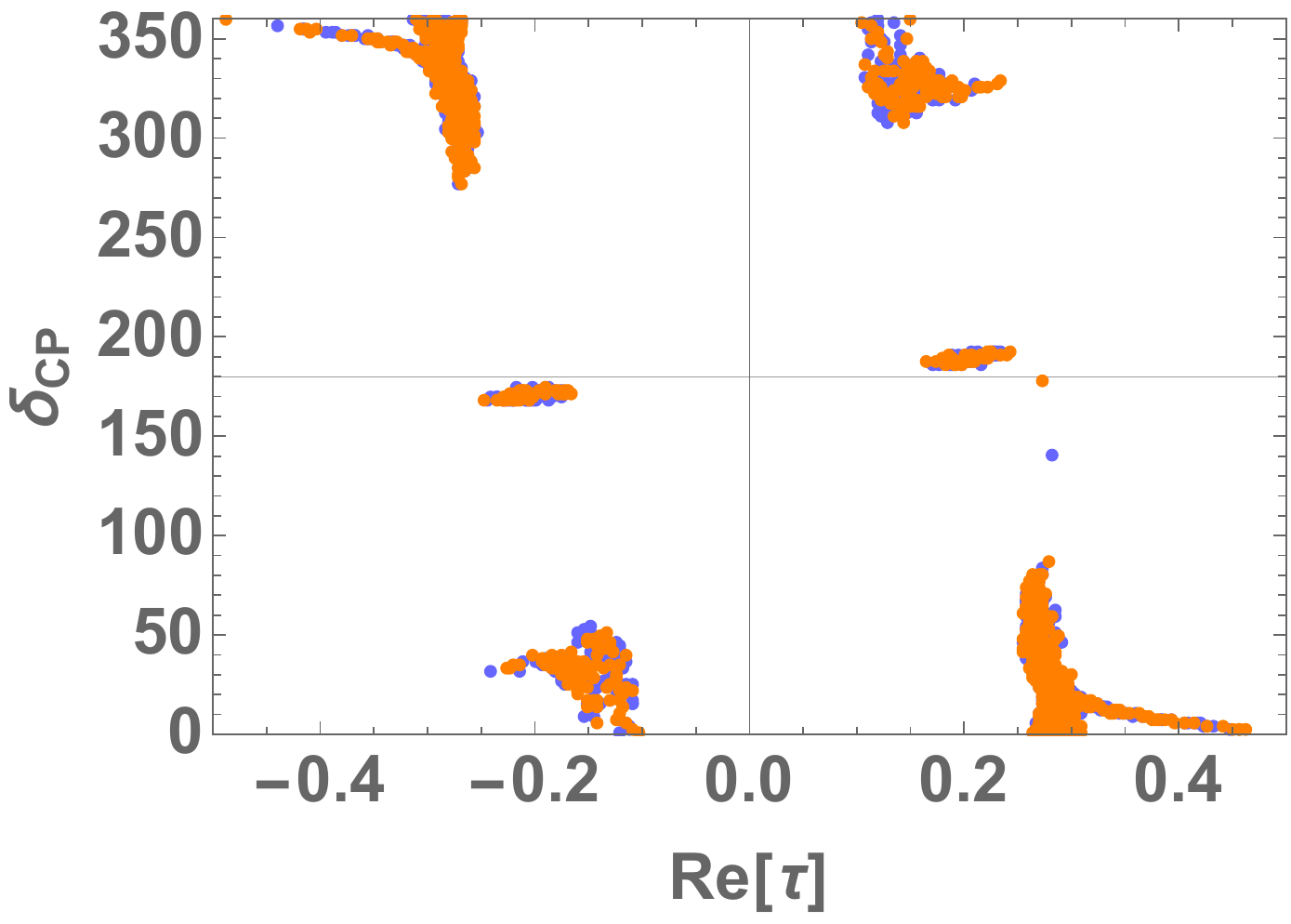}
 		\caption{ Predictive  $\delta_{CP}$ versus ${\rm Re}[\tau]$.
 			Blue color points (negative $Y_B$) almost overlap with orange ones
 			(positive $Y_B$).
 			Points  correspond to the output
 			of section 4 at    $\sqrt{\chi^2}\leq 3$
 		}
 		\label{dpcRetau}
 	\end{minipage}
 \end{figure}

Let us then show the results of the BAU by right-handed neutrinos in our model by using parameter sets 
of section 4 at    $\sqrt{\chi^2}\leq 3$
for NH of neutrino masses.
At first, we discuss the sign of the BAU produced by right-handed neutrinos in the model.   
The  sign of the BAU is determined by the  $\tau$ and the sign of 
the real parameter $g_D$ as shown in Fig.\,\ref{gDRetau}.
In order to obtain the observed positive $Y_B$ (orange points),
the region of 
  (${\rm Re}[\tau]<0$,\,$g_D<0$)  or   (${\rm Re}[\tau]>0$,\,$g_D>0$)
  is required.
  
 In order to see the link between the sign of $Y_B$
  and  the predictive Dirac  CP  phase,
   we show  the predictive $\delta_{CP}$ versus ${\rm Re}[\tau]$
    for the positive (orange) and negative (blue)  $Y_B$ in Fig.\,\ref{dpcRetau}, which is essentially same one in 
    Fig.\,\ref{delcp-reltau} apart from the sign of $Y_B$.
   All six predicted regions of the ${\rm Re}[\tau]$--$\delta_{CP}$
   plane  can give
    both  positive  (orange) and  negative (blue) $Y_B$
     by the choice of the relevant sign of  $g_D$.
 
\begin{figure}[H]
	\begin{minipage}[]{0.47\linewidth}
	\vspace{2mm}
	\includegraphics[{width=\linewidth}]{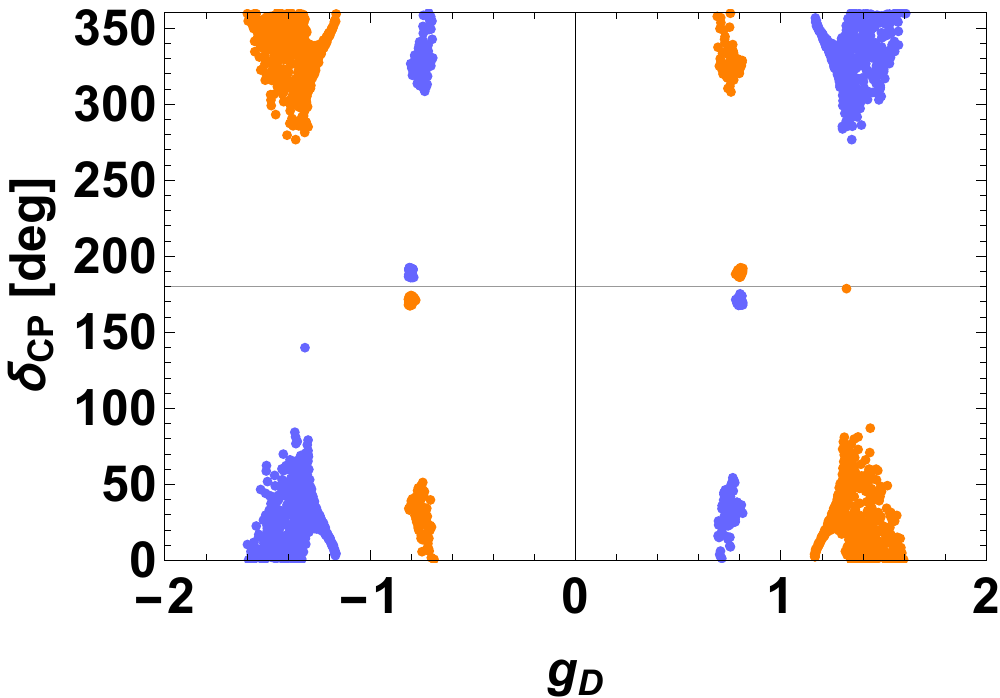}
	\caption{The  sign of $Y_B$ in the  $g_D$--$\delta_{CP}$ plane.
		Orange and blue points
		denote positive $Y_B$ and negative $Y_B$, respectively.
	}
	\label{dcpgD}
\end{minipage}
	\hspace{5mm}
	\begin{minipage}[]{0.47\linewidth}
		\vspace{2mm}
		\includegraphics[{width=\linewidth}]{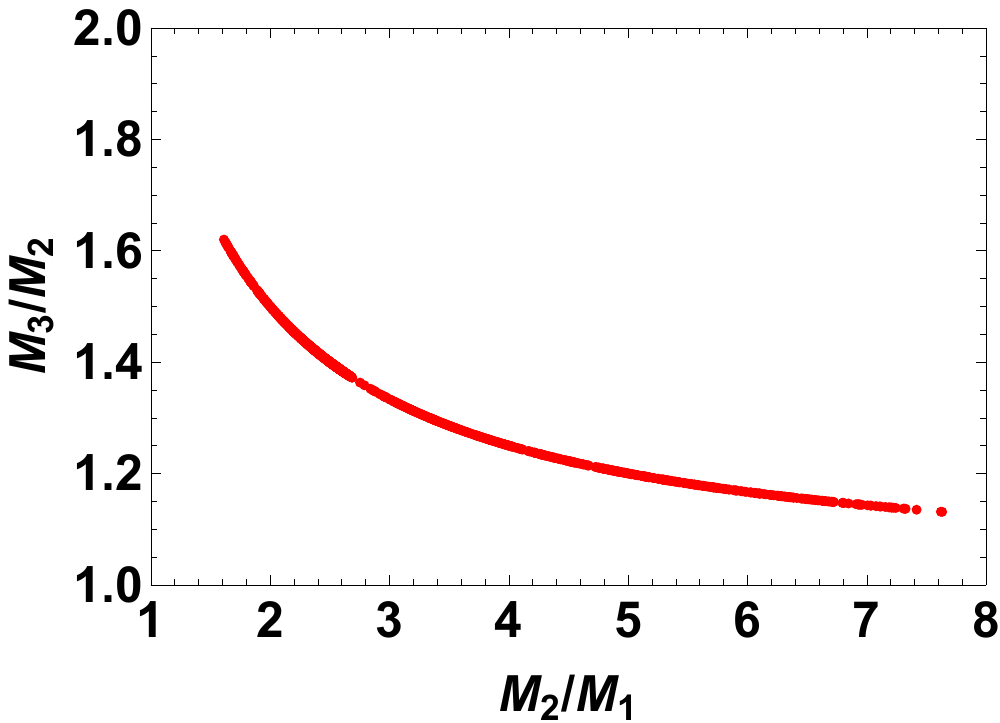}
		\caption{The allowed curve of ratios of right-handed Majorana masses in the 	$M_2/M_1$--$M_3/M_2$ plane.	}
		\label{M123}
	\end{minipage}
\end{figure}

Indeed, we can see this situation  in Fig.\,\ref{dcpgD},
where   the positive (orange) and negative (blue) signs of $Y_B$ are shown   in  the $g_D$--$\delta_{CP}$ plane.
The positive  and negative regions of $Y_B$ are clearly separated.

Next, we discuss the magnitude of the BAU yield.
The yield of the BAU depends on the masses $M_i$ and Yukawa coupling constants of right-handed neutrinos.
These parameters are highly restricted due to the symmetry in our model.
First, the allowed range of the mass ratios of right-handed neutrinos
is shown in Fig.\,\ref{M123}.
It is found that  $M_3/M_2$ decreases  on the curve
 in the range of $[1.1,1.6]$
depending on  $M_2/M_1=[1.6,7.6]$.
Those mass ratios suggest that all three right-handed neutrinos should be taken into account  in the calculation of the leptogenesis.

\begin{figure}[H]
		\begin{minipage}[]{0.47\linewidth}
		\vspace{2mm}
		\includegraphics[{width=\linewidth}]{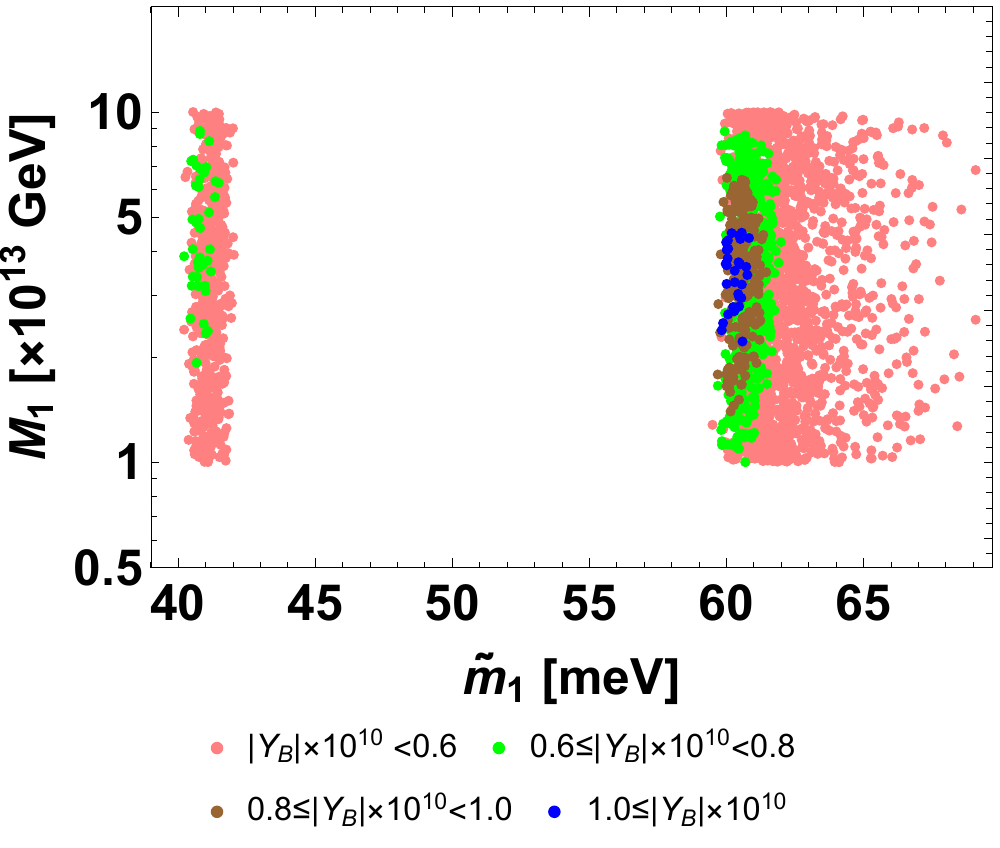}
		\caption{Plot of $\tilde m_1$ and $M_1$ for each $|Y_B|$.	}
		\label{tilde-m1}
	\end{minipage}	
	\hspace{5mm}
\begin{minipage}[]{0.47\linewidth}
	\vspace{2mm}
	\includegraphics[{width=\linewidth}]{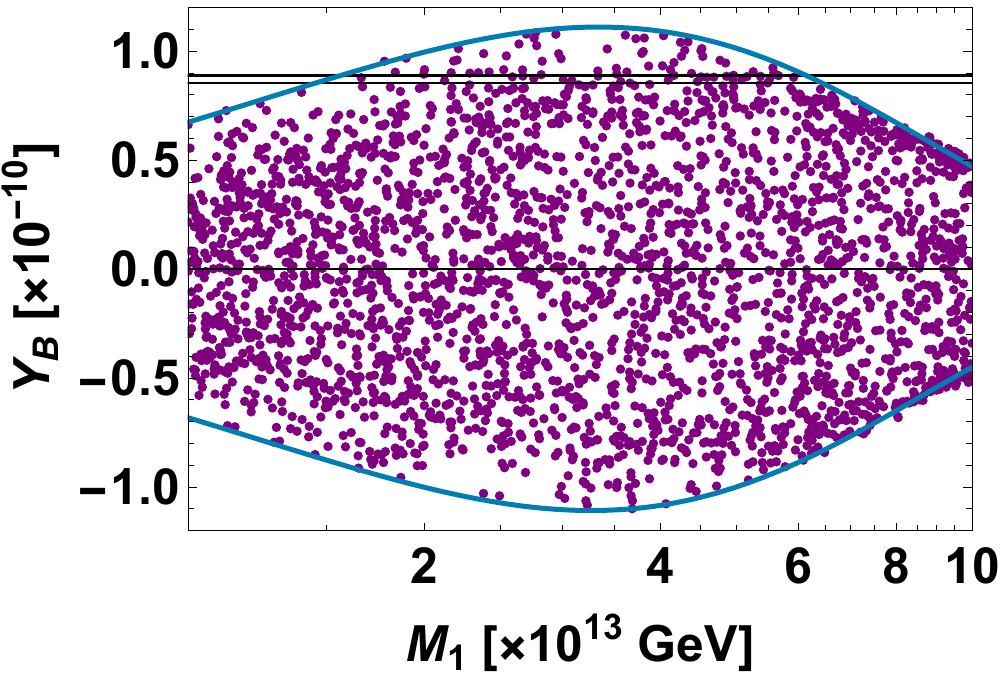}
	\caption{
		Predictive  $Y_B$ versus $M_1$.
		Points  correspond to the  output
		of section 4 at    $\sqrt{\chi^2}\leq 3$.
		Horizontal lines denote the upper and lower bounds
		 of observed $Y_B$ in Eq.\,(\ref{observedYB}).
		The blue solid curves denote 
		 the  boundary of   $Y_B$.
	}
	\label{M1vsYB}
\end{minipage}
\end{figure}

We find that the yield  can be at most the same order of the observed value of the BAU in  Eq.\,(\ref{observedYB}).
This is because the model predicts a relatively large value of
the effective neutrino mass of the leptogenesis $\tilde m_1$ which is defined as $\tilde m_1=(y_{\nu} {y_{\nu}}^{\dagger})_{11}{v_u}^2/M_1$. 
We show the predictive $Y_B$ in the  $\tilde m_1 $--$M_1$ plane
 in Fig.\,\ref{tilde-m1}, where four predictive ranges of $Y_B$
  are discriminated  by colors.
We find numerically 
$\tilde m_1 \simeq 40 \ \rm{meV}$ or $\tilde m_1 \simeq 60 \ \rm{meV}$, and then the strong wash-out effect is inevitable.
In order to obtain the observed BAU, 
$\tilde m_1=[60,\,61 ] \, \rm{meV}$ 
and $M_1 =[1.5,\,6.5]  \times 10^{13}$\,GeV
are  required as seen in Fig.\,\ref{tilde-m1}.
It is an important consequence that the lightest right-handed neutrino 
mass should be in the restricted range.
Thus, the absolute values of right-handed neutrino masses
can be determined from the BAU.

We show predictive $Y_B$ versus $M_1$
 in  Fig.\,\ref{M1vsYB}, where  $M_1$ is taken to be
  ${\cal O} (10^{13})$\,GeV.
The predictive  $Y_B$ is rather broad in this range of $M_1$.
 Especially, it expands maximally at $M_1 = 3.36\times 10^{13}$\,GeV.
That is  because the larger $M_1$ is, the more the wash-out effect of the $\Delta L=2$ processes is important. Thereby, the lightest right-handed neutrino mass is restricted to the specific range
$M_1 =[1.5,\,6.5]  \times 10^{13}$\,GeV.

\begin{figure}[H]
\begin{minipage}[]{0.46\linewidth}
	\vspace{0mm}
	\includegraphics[{width=\linewidth}]
	{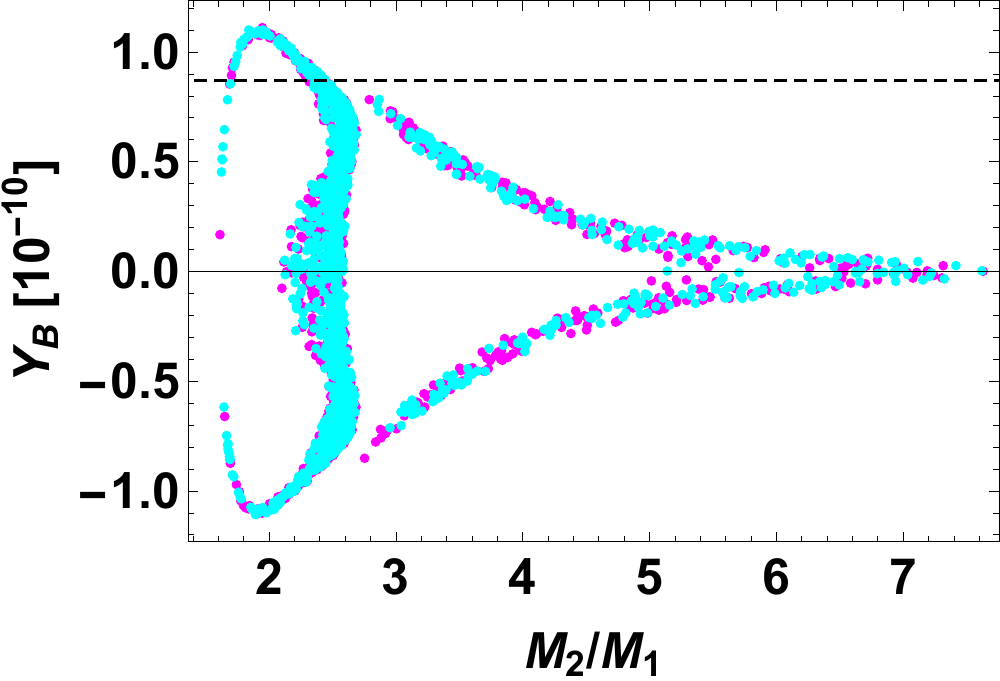}
	\caption{Predictive $Y_B$ versus the mass ratio $M_2/M_1$
		at $M_1 = 3.36 \times 10^{13}$\,GeV.
		Points  correspond to the low energy output
		of section 4 at    $\sqrt{\chi^2}\leq 3$,
		where
		cyan and magenta  correspond to
		positive and negative ${\rm Re}\,[\tau]$,
		respectively.
		Horizontal dashed line denotes the
			central value of observed $Y_B$.	}
	\label{M12ratiovsYB}
\end{minipage}
\hspace{5mm}	
\begin{minipage}[]{0.47\linewidth}
	\vspace{2mm}
	\includegraphics[{width=\linewidth}]{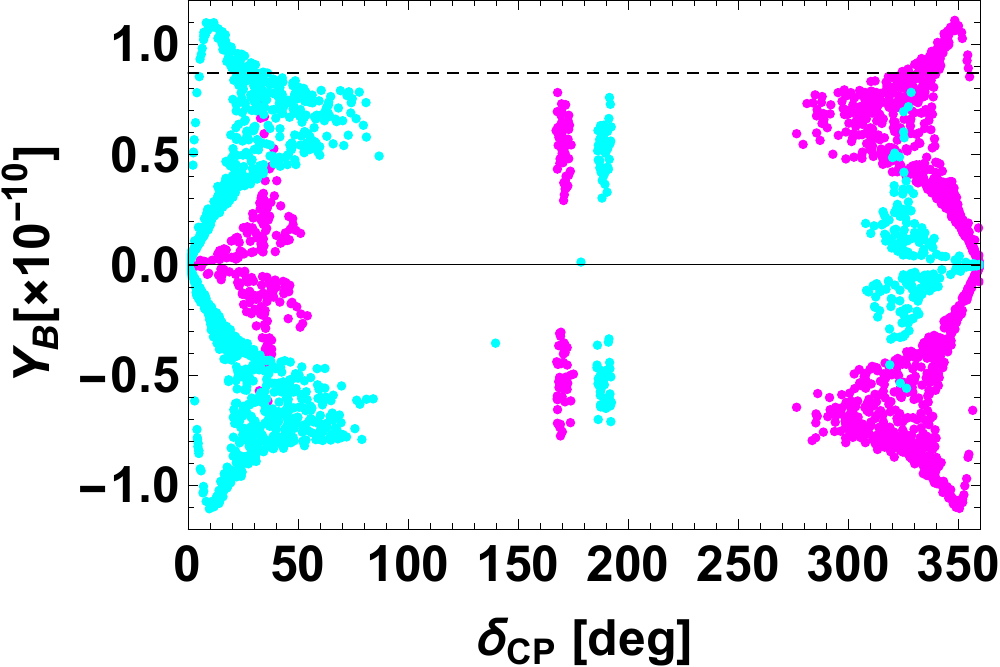}
	\caption{
		Predictive  $Y_B$ versus $\delta_{CP}$ at $M_1 = 3.36 \times 10^{13}$\,GeV.
		Points  correspond to the low energy output
		of section 4 at    $\sqrt{\chi^2}\leq 3$,
		where
		cyan and magenta  correspond to
		positive and negative ${\rm Re}\,[\tau]$,
		respectively.
		Horizontal dashed line denotes the 
		central value of observed $Y_B$.
	}
	\label{delcpvsYB}
\end{minipage}
\end{figure}
In order to see which parameter causes the predictive broad $Y_B$
of  Fig.\,\ref{M1vsYB}, we show the $M_2/M_1$ dependence of    $Y_B$ at $M_1 = 3.36 \times 10^{13}$\,GeV 
in Fig.\,\ref{M12ratiovsYB}.
It is clearly found that the  predictive magnitude of $Y_B$ depends on $M_2/M_1$ crucially in addition  to the magnitude of $M_1$.
If $M_2/M_1$ is fixed in $[1.6,\,2]$, the predictive $Y_B$ 
is in the narrow range, which is consistent with the observed one.
 However, the present neutrino oscillation data still allow
 the  range $M_2/M_1=[1.6,7.6]$ 
 because of the broad  $\tau$  region in Fig.\,\ref{tau}.

Finally,
we present $Y_B$ versus $\delta_{CP}$ 
at $M_1 = 3.36 \times 10^{13}$\,GeV 
to see the correlation
between  the predictive $Y_B$ and the low energy CP violating measure $\delta_{CP}$ in Fig.\,\ref{delcpvsYB}.
As seen in Fig.\,\ref{delcp-reltau}, there are six regions
for the predictive $\delta_{CP}$ versus ${\rm Re}\,[\tau]$. Among them, only two regions are 
available to reproduce the observed BAU. 
As seen in Fig.\,\ref{delcpvsYB}, 
the predictive $\delta_{CP}$ ranges  of  $[0^\circ,50^\circ]$  and $[170^\circ,175^\circ]$ (${\rm Re}\,[\tau]<0$, magenta) cannot reach the observed $Y_B$,
and also  $[185^\circ,190^\circ]$ and  $[310^\circ,360^\circ]$ 
(${\rm Re}\,[\tau]>0$, cyan) 
cannot reach it, but   $[5^\circ,40^\circ]$
(${\rm Re}\,[\tau]>0$, cyan)
and $[320^\circ,355^\circ]$  (${\rm Re}\,[\tau]<0$, magenta)
attain to the observed $Y_B$. 
In these regions, the sum of neutrino masses $\sum m_i$
 are expected to be $[66,\,84]$\,meV,
  which is read from the output in Fig.\,\ref{Dirac1}.
In conclusion,  the precise determination of 
the Dirac CP phase $\delta_{CP}$ and the sum of neutrino masses 
$\sum m_i$  will  test our model in the future.


 We have discussed the leptogenesis  in  the model of
weights ($k_e=-1,\ k_\mu=-3,\ k_\tau=-5$) for right-handed charged leptons.
Actually,  we have also examined the case of 
other alternative weights in Appendix D.
For the case of 
($k_e=-1,\ k_\mu=-1,\ k_\tau=-5$),
 the predictive $Y_B$ is marginal to reproduce the observed one.
 On the other hand,
 the $\mathrm {CP}$ asymmetry parameter  $\varepsilon_I$ of Eq.\,(\ref{asymmetric-parameter})
  is much smaller than the above result
  in the  cases of
 ($k_e=-1,\ k_\mu=-1,\ k_\tau=-7$) and
($k_e=-1,\ k_\mu=-3,\ k_\tau=-7$).
 Thus, the leptogenesis provides a crucial test to select
 the favorable models of leptons.

\section{Summary}

 We have presented $A_4$  modular invariant flavor models of leptons with  the  CP invariance.
The origin of the  CP violation is  only in the modulus $\tau$.
Both CP and modular symmetries are broken spontaneously by the VEV of the modulus $\tau$.  
We have discussed the phenomenological implication of this model, that is flavor mixing angles and CP violating phases.

 We have found  allowed region of $\tau$
  which is consistent with the observed lepton mixing angles and lepton masses
  for NH at    $\sqrt{\chi^2}\leq 3$.
 The CP violating Dirac phase $\delta_{CP}$
  is predicted in 
  $[0^\circ,50^\circ]$, $[170^\circ,175^\circ]$ and
 $[280^\circ,360^\circ]$ for ${\rm Re}\,[\tau]<0$,
 and 
 $[0^\circ,80^\circ]$, $[185^\circ,190^\circ]$ and
 $[310^\circ,360^\circ]$ for ${\rm Re}\,[\tau]>0$.

  The predicted $\sum m_i$ is in  $[60,\,84]$\,meV.
 By using  the predicted Dirac phase and the Majorana phases,
we have obtained the effective mass 
$\langle m_{ee}\rangle$ for the $0\nu\beta\beta$ decay, which is
 in [0.003,\,3]\,meV.
 
 We have also studied the case of IH of neutrino masses.
 There are  no parameter sets which 
 reproduce the observed masses and  three mixing angles 
 below    $\sqrt{\chi^2}= 4$.
 but we have found  parameter sets 
 at    $\sqrt{\chi^2}=4$--$5$.
 
Our CP invariant lepton  mass matrices have  minimum number of parameters, 
eight, apart from the overall scale by putting  weights
 ($k_e=-1,\ k_\mu=-3,\ k_\tau=-5$) for right-handed charged leptons.
 However, this choice for weights
 is not a unique one.
 We have examined alternative three choices:
 ($k_e=-1,\ k_\mu=-1,\ k_\tau=-5$),
 ($k_e=-1,\ k_\mu=-1,\ k_\tau=-7$) and
 ($k_e=-1,\ k_\mu=-3,\ k_\tau=-7$).
 In those three models,  we have also  obtained successful numerical results.

 The modulus $\tau$ links the Dirac CP phase  to the  baryon asymmetry. 
 We have studied the leptogenesis in our model with NH of neutrino masses.
 The sign of the BAU
 is determined by the signs of both ${\rm Re}[\tau]$ and   $g_D$.
 In order to obtain the observed positive $Y_B$, 
 (${\rm Re}[\tau]<0$,\,$g_D<0$) or   (${\rm Re}[\tau]>0$,\,$g_D>0$)
 is required.
 Due to the strong wash-out effect,   the yield  can be at most the same order of the observed value of the BAU.
 Then, the lightest right-handed neutrino 
is in the restricted range
 $M_1 =[1.5,\,6.5]  \times 10^{13}$\,GeV.
In addition, the  predictive $Y_B$ also  depends on $M_2/M_1$ crucially.
 
We have found the correlation
 between  the predictive $Y_B$ and the low energy CP violating measure $\delta_{CP}$.
 Among six regions
 of  the predictive $\delta_{CP}$ versus ${\rm Re}\,[\tau]$,
 only two ranges  of   $[5^\circ,40^\circ]$
 (${\rm Re}\,[\tau]>0$)
 and $[320^\circ,355^\circ]$  (${\rm Re}\,[\tau]<0$)
 are consistent with the BAU, where 
  the sum of neutrino masses $\sum m_i$
 is $[66,\,84]$\,meV. 

  Thus, our scheme of the modulus $\tau$  linking the Dirac CP phase  
  to the baryon asymmetry  gives rise to the idea for  an important test of the phase relevant to leptogenesis.

\section*{Acknowledgments}
This research of H.O. was supported by an appointment to the JRG Program at the APCTP through the Science and Technology Promotion Fund and Lottery Fund of the Korean Government. This was also supported by
 the Korean Local Governments - Gyeongsangbuk-do Province and Pohang City (H.O.). 
 The work of T.Y. was  supported by JSPS KAKENHI Grant Numbers  20H01898.
 H.O. is sincerely grateful for the KIAS member. 
\appendix
\section*{Appendix}

\section{Tensor product of  $A_4$ group}
We take the generators of $A_4$ group for the triplet 
in the symmetric base as follows:
\begin{align}
\begin{aligned}
S=\frac{1}{3}
\begin{pmatrix}
-1 & 2 & 2 \\
2 &-1 & 2 \\
2 & 2 &-1
\end{pmatrix},
\end{aligned}
\qquad 
\begin{aligned}
T=
\begin{pmatrix}
1 & 0& 0 \\
0 &\omega& 0 \\
0 & 0 & \omega^2
\end{pmatrix}, 
\end{aligned}
\label{STbase}
\end{align}
where $\omega=e^{i\frac{2}{3}\pi}$ for a triplet.
In this base,
the multiplication rule is
\begin{align}
\begin{pmatrix}
a_1\\
a_2\\
a_3
\end{pmatrix}_{\bf 3}
\otimes 
\begin{pmatrix}
b_1\\
b_2\\
b_3
\end{pmatrix}_{\bf 3}
&=\left (a_1b_1+a_2b_3+a_3b_2\right )_{\bf 1} 
\oplus \left (a_3b_3+a_1b_2+a_2b_1\right )_{{\bf 1}'} \nonumber \\
& \oplus \left (a_2b_2+a_1b_3+a_3b_1\right )_{{\bf 1}''} \nonumber \\
&\oplus \frac13
\begin{pmatrix}
2a_1b_1-a_2b_3-a_3b_2 \\
2a_3b_3-a_1b_2-a_2b_1 \\
2a_2b_2-a_1b_3-a_3b_1
\end{pmatrix}_{{\bf 3}}
\oplus \frac12
\begin{pmatrix}
a_2b_3-a_3b_2 \\
a_1b_2-a_2b_1 \\
a_3b_1-a_1b_3
\end{pmatrix}_{{\bf 3}\  } \ , \nonumber \\
\nonumber \\
{\bf 1} \otimes {\bf 1} = {\bf 1} \ , \qquad &
{\bf 1'} \otimes {\bf 1'} = {\bf 1''} \ , \qquad
{\bf 1''} \otimes {\bf 1''} = {\bf 1'} \ , \qquad
{\bf 1'} \otimes {\bf 1''} = {\bf 1} \  ,
\end{align}
where
\begin{align}
T({\bf 1')}=\omega\,,\qquad T({\bf 1''})=\omega^2. 
\end{align}
More details are shown in the review~\cite{Ishimori:2010au,Ishimori:2012zz}.

\section{Modular forms in $A_4$ symmetry}
For $\Gamma_3\simeq A_4$, the dimension of the linear space 
${\cal M}_k(\Gamma{(3)})$ 
of modular forms of weight $k$ is $k+1$ \cite{Gunning:1962,Schoeneberg:1974,Koblitz:1984}, i.e., there are three linearly 
independent modular forms of the lowest non-trivial weight $2$.
These forms have been explicitly obtained \cite{Feruglio:2017spp} in terms of
the Dedekind eta-function $\eta(\tau)$: 
\begin{equation}
\eta(\tau) = q^{1/24} \prod_{n =1}^\infty (1-q^n)~, 
\quad\qquad  q= \exp \ (i 2 \pi  \tau )~,
\label{etafunc}
\end{equation}
%
where $\eta(\tau)$ is a  so-called  modular form of weight~$1/2$. 
We  use the  base of the generators $S$ and $T$  in   Eq.\,(\ref{STbase})
for the triplet representation.
Then,
the  modular forms of weight 2 $(k=2)$ transforming
as a triplet of $A_4$, ${ Y^{\rm (2)}_{\bf 3}}(\tau)=(Y_1(\tau),\, Y_2(\tau),\, Y_3(\tau))^T$, can be written in terms of 
$\eta(\tau)$ and its derivative \cite{Feruglio:2017spp}:
\begin{eqnarray} 
\label{eq:Y-A40}
Y_1(\tau) &=& \frac{i}{2\pi}\left( \frac{\eta'(\tau/3)}{\eta(\tau/3)}  +\frac{\eta'((\tau +1)/3)}{\eta((\tau+1)/3)}  
+\frac{\eta'((\tau +2)/3)}{\eta((\tau+2)/3)} - \frac{27\eta'(3\tau)}{\eta(3\tau)}  \right), \nonumber \\
Y_2(\tau) &=& \frac{-i}{\pi}\left( \frac{\eta'(\tau/3)}{\eta(\tau/3)}  +\omega^2\frac{\eta'((\tau +1)/3)}{\eta((\tau+1)/3)}  
+\omega \frac{\eta'((\tau +2)/3)}{\eta((\tau+2)/3)}  \right) , \label{Yi} \\ 
Y_3(\tau) &=& \frac{-i}{\pi}\left( \frac{\eta'(\tau/3)}{\eta(\tau/3)}  +\omega\frac{\eta'((\tau +1)/3)}{\eta((\tau+1)/3)}  
+\omega^2 \frac{\eta'((\tau +2)/3)}{\eta((\tau+2)/3)}  \right)\,.
\nonumber
\end{eqnarray}
%
The overall coefficient in Eq.\,(\ref{Yi}) is 
one possible choice.
It cannot be uniquely determined.
The triplet modular forms of weight $2$
have the following  $q$-expansions:
\begin{align}
{ Y^{\rm (2)}_{\bf 3}}(\tau)
=\begin{pmatrix}Y_1(\tau)\\Y_2(\tau)\\Y_3(\tau)\end{pmatrix}=
\begin{pmatrix}
1+12q+36q^2+12q^3+\dots \\
-6q^{1/3}(1+7q+8q^2+\dots) \\
-18q^{2/3}(1+2q+5q^2+\dots)\end{pmatrix}.
\label{Y(2)}
\end{align}
%
They satisfy also the constraint \cite{Feruglio:2017spp}:
\begin{align}
Y_2(\tau)^2+2Y_1(\tau) Y_3(\tau)=0~.
\label{condition}
\end{align}

The  modular forms of the  higher weight, $k$, can be obtained
by using the $A_4$ tensor products of Appendix A 
in terms of   the modular forms  with weight 2,
${ Y^{\rm (2)}_{\bf 3}}(\tau)$. 

\section{Majorana and Dirac phases and $\langle m_{ee}\rangle $
	in  $0\nu\beta\beta$ decay }

Supposing neutrinos to be Majorana particles, 
the PMNS matrix $U_{\text{PMNS}}$~\cite{Maki:1962mu,Pontecorvo:1967fh} 
is parametrized in terms of the three mixing angles $\theta _{ij}$ $(i,j=1,2,3;~i<j)$,
one CP violating Dirac phase $\delta _\text{CP}$ and two Majorana phases 
$\alpha_{21}$, $\alpha_{31}$  as follows:
\begin{align}
U_\text{PMNS} =
\begin{pmatrix}
c_{12} c_{13} & s_{12} c_{13} & s_{13}e^{-i\delta_\text{CP}} \\
-s_{12} c_{23} - c_{12} s_{23} s_{13}e^{i\delta_\text{CP}} &
c_{12} c_{23} - s_{12} s_{23} s_{13}e^{i\delta_\text{CP}} & s_{23} c_{13} \\
s_{12} s_{23} - c_{12} c_{23} s_{13}e^{i\delta_\text{CP}} &
-c_{12} s_{23} - s_{12} c_{23} s_{13}e^{i\delta_\text{CP}} & c_{23} c_{13}
\end{pmatrix}
\begin{pmatrix}
1&0 &0 \\
0 & e^{i\frac{\alpha_{21}}{2}} & 0 \\
0 & 0 & e^{i\frac{\alpha_{31}}{2}}
\end{pmatrix},
\label{UPMNS}
\end{align}
where $c_{ij}$ and $s_{ij}$ denote $\cos\theta_{ij}$ and $\sin\theta_{ij}$, respectively.

The rephasing invariant CP violating measure of leptons \cite{Jarlskog:1985ht,Krastev:1988yu}
is defined by the PMNS matrix elements $U_{\alpha i}$. 
It is written in terms of the mixing angles and the CP violating phase as:
\begin{equation}
J_{CP}=\text{Im}\left [U_{e1}U_{\mu 2}U_{e2}^\ast U_{\mu 1}^\ast \right ]
=s_{23}c_{23}s_{12}c_{12}s_{13}c_{13}^2\sin \delta_\text{CP}\,,
\label{Jcp}
\end{equation}
where $U_{\alpha i}$ denotes the each component of the PMNS matrix.

There are also other invariants $I_1$ and $I_2$ associated with Majorana phases
\begin{equation}
I_1=\text{Im}\left [U_{e1}^\ast U_{e2} \right ]
=c_{12}s_{12}c_{13}^2\sin \left (\frac{\alpha_{21}}{2}\right )\,, \quad
I_2=\text{Im}\left [U_{e1}^\ast U_{e3} \right ]
=c_{12}s_{13}c_{13}\sin \left (\frac{\alpha_{31}}{2}-\delta_\text{CP}\right )\,.
\label{Jcp}
\end{equation}
We can calculate $\delta_\text{CP}$, $\alpha_{21}$ and $\alpha_{31}$ with these relations by taking account of 
\begin{eqnarray}
&&\cos\delta_{CP}=\frac{|U_{\tau 1}|^2-
	s_{12}^2 s_{23}^2 -c_{12}^2c_{23}^2s_{13}^2}
{2 c_{12}s_{12}c_{23}s_{23}s_{13}}\,, \nonumber \\
&&\text{Re}\left [U_{e1}^\ast U_{e2} \right ]
=c_{12}s_{12}c_{13}^2\cos \left (\frac{\alpha_{21}}{2}\right )\,, \qquad
\text{Re}\left [U_{e1}^\ast U_{e3} \right ]
=c_{12}s_{13}c_{13}\cos\left(\frac{\alpha_{31}}{2}-\delta_\text{CP}\right )\,.
\end{eqnarray}
In terms of this parametrization, the effective mass for the $0\nu\beta\beta$ decay is given as follows:
\begin{align}
\langle m_{ee}	\rangle=\left| m_1 c_{12}^2 c_{13}^2+ m_2s_{12}^2 c_{13}^2 e^{i\alpha_{21}}+
m_3 s_{13}^2 e^{i(\alpha_{31}-2\delta_{CP})}\right|  \, .
\end{align}


\section{Alternative $A_4$ modular models}
We present three alternative charged lepton mass matrices 
 by putting different weights for ($k_e,\ k_\mu,\ k_\tau$),
which are consistent with observed mixing angles and masses.
Neutrino mass matrix is the same one in section \ref{A4model}.

\subsection{Case of ($k_e=-1,\ k_\mu=-1,\ k_\tau=-5 $) for charged leptons}
The assignment of MSSM fields and modular forms are given 
in Table \ref{tb:lepton226}.
\begin{table}[H]
	\centering
	\begin{tabular}{|c||c|c|c|c|c|c|} \hline
		\rule[14pt]{0pt}{1pt}
		&$L$&$(e^c,\mu^c,\tau^c)$&$N^c$ &$H_u$&$H_d$&$ Y_{\bf 3}^{\rm (2)}, 
		\ \      Y_{\bf r}^{\rm (6)}$ \\  \hline\hline 
		\rule[14pt]{0pt}{1pt}
		$SU(2)$&$\bf 2$&$\bf 1$& $\bf 1$ &$\bf 2$&$\bf 2$&$\bf 1$\\
		\rule[14pt]{0pt}{1pt}
		$A_4$&$\bf 3$& \bf (1,\ 1$''$,\ 1$'$)& $\bf 3$ &$\bf 1$&$\bf 1$ & 
	\quad	$\bf 3,\quad   \{3, 3'\}$\\
		\rule[14pt]{0pt}{1pt}
		$k$ & $ -1$ & $(-1,\ -3,\ -5)$ & $-1$ & $0$ & $0$ & \hskip -0. cm $2, \qquad\  6$ \\ \hline
	\end{tabular}	
	\caption{ Representations and  weights
		$k$ for MSSM fields and  modular forms of weights $2$ and $6$.
	}
	\label{tb:lepton226}
\end{table}
The $A_4$ invariant superpotential of the charged leptons, $w_E$,
by taking into account the modular weights is obtained as 
\begin{align}
w_E=\alpha_e e^c H_d { Y^{\rm (2)}_{\bf 3}}L+
\beta_e \mu^c H_d { Y^{\rm (4)}_{\bf 3}}L+
\gamma_e \tau^c H_d { Y^{\rm (6)}_{\bf 3}}L+
\gamma_e' \tau^c H_d { Y^{\rm (6)}_{\bf 3'}}L\, . 
\end{align}
By using  $g_e=\gamma _e '/\gamma _e $, the charged lepton mass matrix $M_E$  is simply written  as:    
\begin{align}
\begin{aligned}
M_E(\tau)=v_d \begin{pmatrix}
\alpha_e & 0 & 0 \\
0 &\beta_e & 0\\
0 & 0 &\gamma_e
\end{pmatrix}
\begin{pmatrix}
Y_1(\tau) & Y_3(\tau) & Y_2(\tau) \\
Y_2(\tau) & Y_1(\tau) & Y_3(\tau) \\
Y_3^{(6)}(\tau)+g_eY_3'^{(6)}(\tau) & Y_2^{(6)}(\tau)+g_eY_2'^{(6)}(\tau) & 
Y_1^{(6)}(\tau)+g_eY_1'^{(6)}(\tau)
\end{pmatrix}_{RL} \ .
\end{aligned}
\label{ME(2)246}
\end{align}

\subsection{Case of ($k_e=-1,\ k_\mu=-1,\ k_\tau=-7 $) for charged leptons}
The assignment of MSSM fields and modular forms are given 
in Table \ref{tb:lepton228}.
\begin{table}[H]
	\centering
	\begin{tabular}{|c||c|c|c|c|c|c|} \hline
		\rule[14pt]{0pt}{1pt}
	&$L$&$(e^c,\mu^c,\tau^c)$&$N^c$ &$H_u$&$H_d$&$ Y_{\bf 3}^{\rm (2)}, 
		\ \   Y_{\bf r}^{\rm (8)}$\\  \hline\hline 
		\rule[14pt]{0pt}{1pt}
		$SU(2)$&$\bf 2$&$\bf 1$& $\bf 1$ &$\bf 2$&$\bf 2$&$\bf 1$\\
		\rule[14pt]{0pt}{1pt}
		$A_4$&$\bf 3$& \bf (1,\ 1$''$,\ 1$'$)& $\bf 3$ &$\bf 1$&$\bf 1$&$\bf 3, \ \ \ \{3, 3'\}$\\
		\rule[14pt]{0pt}{1pt}
		$k$ & $ -1$ & $(-1,\ -1,\ -7)$ & $-1$ & $0$ & $0$ & \hskip -0.3 cm $2, \quad 8$ \\ \hline
	\end{tabular}	
	\caption{ Representations and  weights
		$k$ for MSSM fields and  modular forms of weights $2$ and $8$.
	}
	\label{tb:lepton228}
\end{table}
For $k=8$, there are  9 modular forms
by the tensor products of  $\rm A_4$ as:
\begin{align}
&\begin{aligned}
{ Y^{(\rm 8)}_{\bf 1}}=(Y_1^2+ 2 Y_2 Y_3)^2 \, ,\qquad  
{ Y^{(\rm 8)}_{\bf 1'}}=(Y_1^2+ 2 Y_2 Y_3)(Y_3^2+ 2 Y_1 Y_2)\, ,  \qquad
{ Y^{(\rm 8)}_{\bf 1"}}=(Y_3^2+ 2 Y_1 Y_2)^2 \, ,
\end{aligned} \nonumber \\
&\begin{aligned} { Y^{(\rm 8)}_{\bf 3}}\equiv 
\begin{pmatrix}
Y_1^{(8)}  \\
Y_2^{(8)} \\
Y_3^{(8)}
\end{pmatrix}
= (Y_1^3+Y_2^3+Y_3^3-3 Y_1 Y_2 Y_3 )
\begin{pmatrix}
Y_1 \\
Y_2 \\
Y_3
\end{pmatrix} , \ 
\end{aligned}
\begin{aligned} { Y^{(\rm 8)}_{\bf 3'}}\equiv
\begin{pmatrix}
Y_1^{'(8)}  \\
Y_2^{'(8)} \\
Y_3^{'(8)}
\end{pmatrix}
=(Y_3^2+ 2 Y_1 Y_2)
\begin{pmatrix}
Y_2^2 - Y_1  Y_3   \\
Y_1^2 - Y_2  Y_3  \\
Y_3^2 - Y_1  Y_2 
\end{pmatrix} . \nonumber
\end{aligned}
\label{weight8}
\end{align}
The $A_4$ invariant superpotential of the charged leptons, $w_E$,
by taking into account the modular weights is obtained as 
\begin{align}
w_E=\alpha_e e^c H_d { Y^{\rm (2)}_{\bf 3}}L+
\beta_e \mu^c H_d { Y^{\rm (2)}_{\bf 3}}L+
\gamma_e \tau^c H_d { Y^{\rm (8)}_{\bf 3}}L+
\gamma_e' \tau^c H_d { Y^{\rm (8)}_{\bf 3'}}L\, . 
\end{align}
By using  $g_e=\gamma _e '/\gamma _e $, the charged lepton mass matrix $M_E$  is simply written  as:   
\begin{align}
\begin{aligned}
M_E(\tau)=v_d \begin{pmatrix}
\alpha_e & 0 & 0 \\
0 &\beta_e & 0\\
0 & 0 &\gamma_e
\end{pmatrix}
\begin{pmatrix}
Y_1(\tau) & Y_3(\tau) & Y_2(\tau) \\
Y_2(\tau) & Y_1(\tau) & Y_3(\tau) \\
Y_3^{(8)}(\tau)+g_eY_3'^{(8)}(\tau) & Y_2^{(8)}(\tau)+g_eY_2'^{(8)}(\tau) & 
Y_1^{(8)}(\tau)+g_eY_1'^{(8)}(\tau)
\end{pmatrix}_{RL} \, .
\end{aligned}
\label{ME(2)228}
\end{align}

\subsection{Case of ($k_e=-1,\ k_\mu=-3,\ k_\tau=-7 $) for charged leptons}
The assignment of MSSM fields and modular forms are given 
in Table \ref{tb:lepton248}.
\begin{table}[H]
	\centering
	\begin{tabular}{|c||c|c|c|c|c|c|} \hline
		\rule[14pt]{0pt}{1pt}
 &$L$&$(e^c,\mu^c,\tau^c)$&$N^c$ &$H_u$&$H_d$&$ Y_{\bf 3}^{\rm (2)}, 
		\ \   Y_{\bf 3}^{\rm (4)},\ \   Y_{\bf r}^{\rm (8)}$ \\  \hline\hline 
		\rule[14pt]{0pt}{1pt}
		$SU(2)$&$\bf 2$&$\bf 1$& $\bf 1$ &$\bf 2$&$\bf 2$&$\bf 1$\\
		\rule[14pt]{0pt}{1pt}
		$A_4$&$\bf 3$& \bf (1,\ 1$''$,\ 1$'$)& $\bf 3$ &$\bf 1$&$\bf 1$ & 
		$\bf 3,\quad \bf 3,\quad  \{3, 3'\}$\\
		\rule[14pt]{0pt}{1pt}
		$k$ & $ -1$ & $(-1,\ -3,\ -7)$ & $-1$ & $0$ & $0$ & \hskip -0.7 cm $2, \quad 4,\ \ \quad 8$ \\ \hline
	\end{tabular}	
	\caption{ Representations and  weights
		$k$ for MSSM fields and  modular forms of weights $2$, $4$ and $8$.
	}
	\label{tb:lepton248}
\end{table}
The $A_4$ invariant superpotential of the charged leptons, $w_E$,
by taking into account the modular weights is obtained as 
\begin{align}
w_E=\alpha_e e^c H_d { Y^{\rm (2)}_{\bf 3}}L+
\beta_e \mu^c H_d { Y^{\rm (4)}_{\bf 3}}L+
\gamma_e \tau^c H_d { Y^{\rm (8)}_{\bf 3}}L+
\gamma_e' \tau^c H_d { Y^{\rm (8)}_{\bf 3'}}L\, .
\end{align}
By using  $g_e=\gamma _e '/\gamma _e $, the charged lepton mass matrix $M_E$  is simply written  as:  
\begin{align}
\begin{aligned}
M_E(\tau)=v_d \begin{pmatrix}
\alpha_e & 0 & 0 \\
0 &\beta_e & 0\\
0 & 0 &\gamma_e
\end{pmatrix}
\begin{pmatrix}
Y_1(\tau) & Y_3(\tau) & Y_2(\tau) \\
Y_2^{(4)}(\tau) & Y_1^{(4)}(\tau) & Y_3^{(4)}(\tau) \\
Y_3^{(8)}(\tau)+g_eY_3'^{(8)}(\tau) & Y_2^{(8)}(\tau)+g_eY_2'^{(8)}(\tau) & 
Y_1^{(8)}(\tau)+g_eY_1'^{(8)}(\tau)
\end{pmatrix}_{RL} \, .
\end{aligned}
\label{ME(2)248}
\end{align}

\subsection{Sample parameters of alternative models}

In Table \ref{sample3}, we show parameters and output of our calculations in above  three cases	 of ($k_e,\ k_\mu,\ k_\tau$) for NH.

\begin{table}[H]
	\centering
	\begin{tabular}{|c|c|c|c|} \hline 
		\rule[14pt]{0pt}{0pt}
($k_e,\ k_\mu,\ k_\tau$)&($-1,\ -1,\ -5$) &($-1,\ -1,\ -7$)&($-1,\ -3,\ -7$)\\  \hline
		\rule[14pt]{0pt}{0pt}
	$\tau$&  $  -0.1912 + 1.1194   \, i$ & $0.0901 + 1.0047 \, i$&
	 -0.1027 + 1.0050\, i \\ 
		\rule[14pt]{0pt}{0pt}
		$g_D$ &$ -0.800$ & $-0.660$& $0.685$\\
		\rule[14pt]{0pt}{0pt}
		$g_e$  &  $-0.905$& $-0.530$& $-0.573$\\
		\rule[14pt]{0pt}{0pt}
		$\beta_e/\alpha_e$ & $3.70\times 10^{-3}$  &$5.94\times 10^{-3}$&
		$6.30\times 10^{-3}$ \\
		\rule[14pt]{0pt}{0pt} 
		$\gamma_e/\alpha_e$ &  $ 9.71$  &$17.6$&$16.0$ \\
		\rule[14pt]{0pt}{0pt}
		$\sin^2\theta_{12}$ & $ 0.305$& $0.324$& $0.326$\\
		\rule[14pt]{0pt}{0pt}
		$\sin^2\theta_{23}$ &  $ 0.569$& $0.441$& $0.479$\\
		\rule[14pt]{0pt}{0pt}
		$\sin^2\theta_{13}$ &  $ 0.0222$&$0.0222$&$0.0223$	\\
		\rule[14pt]{0pt}{0pt}
		$\delta_{CP}^\ell$ &  $172^\circ$& $183^\circ$& $176^\circ$\\
		\rule[14pt]{0pt}{0pt}
		$[\alpha_{21},\,\alpha_{31}]$ &  $[192^\circ,\,263^\circ]$& 
		$[181^\circ,\,-0.1^\circ]$&	$[179^\circ,\,0.2^\circ]$	\\	
		\rule[14pt]{0pt}{0pt}
		$\sum m_i$ &  $62.5$\,meV &	 $60.5$\,meV & $60.7$\,meV \\
		\rule[14pt]{0pt}{0pt}
	$\langle m_{ee} \rangle$ & $1.69$\,meV& $0.58$\,meV & $0.52$\,meV\\
		\rule[14pt]{0pt}{0pt}
		$\sqrt{\chi^2}$ & $1.08$  & $2.16$ & $2.38$\\
		\hline
	\end{tabular}
	\caption{Numerical values of parameters and observables
		at  sample points of NH.}
	\label{sample3}
\end{table}


\section{Formulae of the leptogenesis}

We solve the Boltzmann equations for right-handed neutrinos number densities
$n_{N_I}$ and the lepton asymmetry density $n_L$ as:
\begin{align}
\frac{d Y_{N_I}}{dz}	
&=	
\frac{- z}{s H(M_1)}	\Bigg\{	
\left(	\frac{Y_{N_I}}{Y_{N_I}^{eq}}	-	1	\right) 	
\left(\gamma_{N_I}+2 \gamma_{tI}^{(3)}	+4\gamma_{tI}^{(4)} \right)	
+
\sum_{J=1}^3	
\left( \frac{Y_{N_I}}{Y_{N_I}^{eq}}\frac{Y_{N_J}}{Y_{N_J}^{eq}}	-	1	\right)		
\left( \gamma_{N_I N_J}^{(2)}+\gamma_{N_I N_J}^{(3)} \right)
\Bigg\}	\,,
\label{eq:yieldNI}
\\
\frac{d Y_{L}}{dz}	
&=	
\frac{-z}{s H(M_1)}	\Bigg\{
\sum_{I=1}^3 \left[	
\left( 1 - \frac{Y_{N_I}}{Y_{N_I}^{eq}} \right) \varepsilon_I \, \gamma_{N_I} 
+ \frac{Y_{L}}{Y_{\ell}^{eq}} \frac{\gamma_{N_I}}{2}
\right]	
+	
\frac{Y_{L}}{Y_{\ell}^{eq}}	\left( 2 \gamma_{N}^{(2)}	+ 2\gamma_{N}^{(13)} \right)
\nonumber \\
&~~~~~~~~~+	
\frac{Y_{L}}{Y_{\ell}^{eq}}	
\sum_{I=1}^3
\left[ \frac{Y_{N_I}}{Y_{N_I}^{eq}}\gamma_{tI}^{(3)} 
+ 2\gamma_ {tI}^{(4)}
+ \frac{Y_{N_I}}{Y_{N_I}^{eq}} 
\left( \gamma_{WI}^{(1)} +\gamma_{BI}^{(1)} \right)
+
\gamma_{WI}^{(2)} +\gamma_{WI}^{(3)}+ \gamma_{BI}^{(2)} +\gamma_{BI}^{(3)} 
\right]
\Bigg\}	\,,
\label{eq:yieldYL2}
\end{align}
where $z =M_1/T$. 
Here we define the yields as $Y_{N_I} = n_{N_I}/s$ and $Y_{L} = n_L /s$, 
with the entropy density of the universe $s$. 
The superscript "$eq$" denotes its equilibrium value. We apply the Boltzmann approximation and the yield for a massless particle with one degree of freedom in equilibrium is given by $Y_{\ell}^{eq} = 45/(2\pi^4 g_{\ast s})$, with $g_{\ast s}=110.75$. 

The flavor summed CP asymmetry at the decay of the right-handed neutrino $N_I$ is given as 
\begin{equation}
\varepsilon_{I}=-\frac{1}{8\pi} \sum_{J \neq I}
\frac{{\rm Im}[\{(y_\nu y_\nu^\dagger)_{IJ}\}^2]}
{(y_\nu y_\nu^\dagger)_{II}}
\left [f^V \left (\frac{M_J^2}{M_I^2}\right )+
f^S \left (\frac{M_J^2}{M_I^2}\right )\right ] \ ,
\label{asym}
\end{equation}
where $f^V(x)$ and  $f^S(x)$  are the contributions from vertex and self-energy corrections, respectively.
In the case of the standard model (SM) with right-handed neutrinos,
they are given as
\begin{equation}
f^V(x)=\sqrt{x}\left [ (x+1)\ln\left ( 1+\frac{1}{x}\right )-1\right ], \quad\quad
f^S(x)=\frac{\sqrt{x}}{x-1} .
\end{equation}

The reaction density 
for the $N_I$ decay is given by
\begin{align}
&\gamma_{N_I} 
= \frac{\left(y_\nu {y_\nu}^\dagger \right)_{I I}}{8 \pi^{3}} 
M_{1}^{4} a_{I}^{3/2}
\frac{K_{1}\left(\sqrt{a_{I}} z\right)}{z} \,,
\end{align}
where $z =M_1/T$, $a_I = (M_I / M_1)^2$, and $K_1(x)$ is the modified Bessel function
of the second kind.
Note that $y_\nu$ is the Yukawa coupling matrix of neutrinos in the base where
both the mass matrices of charged leptons and right-handed neutrinos are diagonalized.
The reaction density for the scattering process $A + B \to C + D$ is expressed as
\begin{align}
&\gamma(A+B \rightarrow C+D) 
= \frac{T}{64 \pi^{4}}
\int_{ \left(m_{A}+m_{B}\right)^{2} }^{\infty} 
ds \, 
\hat{\sigma}(s) \sqrt{s} K_{1}\left(\frac{\sqrt{s}}{T}\right)\,,
\end{align} 
where $m_{A}$ and $m_B$ are masses of the initial particles 
and $\hat \sigma (s)$ denotes the reduced cross section for the process.
The expressions of the reduced cross sections for the $\Delta L=1$ processes induced through top Yukawa intraction, the $\Delta L=2$ scattering processes and 
the annihilation processes of right-handed neutrinos are found in Ref.~\cite{Plumacher:1998ex}. 
The reduced cross section for $L \overline{H_u} \to \overline{L}  H_u$~process which is correctly subtracted $N_I$ on-shell contribution is~\cite{Giudice:2003jh}
\begin{align}
\hat{\sigma}_{N}^{(2)}(x)	
&=	
\frac{1}{2\pi} 
\Bigg[ \sum_{I} (y_{\nu} {y_{\nu}}^\dagger)_{II}^2	\frac{a_I}{x} 
\Big\{\frac{x}{a_I}+\frac{x}{D_{I}}	
- \Big(1+\frac{x+a_I}{D_{I}} \Big)
\log \Big(	\frac{x+a_I}{a_I}\Big)	
\Big\}	
\nonumber \\
&~~~
+\sum_{I>J}	\mathrm{Re}[(y_{\nu} {y_{\nu}}^\dagger)_{IJ}^2]
\frac{\sqrt{a_I a_J}}{x} 
\Big\{ 	
\frac{x^2 + x ( D_I + D_J )}{D_{I} D_{J}} 
+ 
(x+a_I) \Big(\frac{2}{a_J-a_I}-\frac{1}{D_J}\Big) 
\ln\Big(\frac{x+a_I}{a_I}\Big)
\nonumber 	\\
&\hspace{5cm}
+
(x+a_J)
\Big(\frac{2}{a_I-a_J}-\frac{1}{D_I} \Big) 
\ln \Big( \frac{x+a_J}{a_J} \Big)	\Big\} \Bigg]	\,,
\end{align}
where $D_I = [(x-a_I)^2 + a_I c_I]/(x-a_I)$ with $c_I = (\Gamma_{N_I}/M_1)^2$, in which $\Gamma_{N_I}$ is the total decay rate of right-handed neutrino $N_I$.
The explicit form of reduced cross sections for $\Delta L=1$ processes through 
the $SU(2)_L$ SM gauge interaction are found in Refs.~\cite{Giudice:2003jh,Pilaftsis:2003gt}, 
\begin{align}
\hat \sigma_{WI}^{(1)}(x)	
&=	
\frac{3g_2^2(y_{\nu} {y_{\nu}}^\dagger)_{II}}{16\pi x^2}
\Big[ -2x^2+6a_I x -4a_I^2	+(x^2-2a_I x +2a_I^2) 
\ln \Big|\frac{x-a_I+a_L}{a_L} \Big|
\nonumber \\
&\hspace{3cm}	
+
\frac{x(a_L x +a_L a_I -a_W a_I)(a_I -x)}{a_L(x-a_I+a_L)}
\Big] \,,
\\
\hat \sigma_{WI}^{(2)}(x)	
&=
\frac{3g_2^2(y_{\nu} {y_{\nu}}^\dagger)_{II}}{8\pi x(x-a_I)}
\Big[
2a_I x \ln \Big| \frac{x -a_I +a_H}{a_H} \Big|	
+ (x^2 +a_I^2) \ln \Big| \frac{x-a_I-a_W-a_H}{-a_W -a_H} \Big|
\Big] \,,
\\
\hat \sigma_{WI}^{(3)}(x)	
&=	
\frac{3g_2^2(y_{\nu} {y_{\nu}}^\dagger)_{II} a_I}{16\pi x^2}
\Big[	\frac{x^2 -4a_I x +3a_I^2}{a_I} 
+ 4(x-a_I) \ln \Big| \frac{x-a_I +a_H}{a_H} \Big|	
- \frac{x(4a_H -a_W)(x-a_I)}{a_H(x-a_I+m_H)}	
\Big] \,.
\end{align}
Here $\hat \sigma_{WI}^{(1)}$, $\hat \sigma_{WI}^{(2)}$ and $\hat \sigma_{WI}^{(3)}$ correspond to the reduced cross sections of the processes $N_I L \rightarrow H_u W$, $N_I W \rightarrow \overline{L} H_u$ and $N_I \overline{H_u} \rightarrow \overline{L} W$, respectively.
We have used $a_{X} = m_{X}^2/M_1^2$
where $m_X$ with $X = L, H_u, W, B$ are thermal masses of lepton doublets, up-type Higgs, $SU(2)_L$ gauge bosons and $U(1)_Y$ gauge boson, respectively.
The reaction densities for the $\Delta L = 1$ processes
through $U(1)_Y$ gauge interaction $\hat \sigma_{BI}^{(i)}$ are obtained by replacing 
$a_W$ with $a_B$ and $\frac{3}{2} g_2^2$ with $\frac{1}{4} g_Y^2$
in $\hat \sigma_{WI}^{(i)}$.  

For the more accurate estimation of the baryon asymmetry, we have taken into account
the one-loop RGE evolutions of couplings and the renormalization scale
is taken as $\mu = 2\pi T$.


\end{document}